\documentclass[11pt]{article}

\usepackage{jheppub} % for details on the use of the package, please
\usepackage{amsmath,amssymb}
\usepackage{url}

\usepackage{graphicx}
\usepackage[font=small,labelfont=bf]{caption}
\usepackage{subcaption}
\usepackage{tikz}
\usepackage[compat=1.1.0]{tikz-feynman}
\usepackage{siunitx}
\usepackage{float}
\usepackage{soul}

\usepackage{color}

\def\Z{\mathbb{Z}}

\title{Topological Freeze-out by Semi-Annihilation}

\author[a]{Joe Davighi,}\emailAdd{joseph.davighi@cern.ch}
\author[b,c]{Serah Moldovsky,}\emailAdd{serah\_moldovsky@berkeley.edu}
\author[b,c,d,1]{Hitoshi Murayama,\note{Hamamatsu Professor}}\emailAdd{hitoshi@berkeley.edu}
\author[b,c]{Christiane Scherb,}\emailAdd{cscherb@lbl.gov}
\author[e]{Nud\v{z}eim Selimovi\'c}\emailAdd{nudzeim.selimovic@pd.infn.it}

\affiliation[a]{Theoretical Physics Department, CERN, 1211 Geneva 23, Switzerland}
\affiliation[b]{Department of Physics, University of California, Berkeley, CA 94720, USA}
\affiliation[c]{Ernest Orlando Lawrence Berkeley National Laboratory, Berkeley, CA 94720, USA}
\affiliation[d]{Kavli Institute for the Physics and Mathematics of the Universe (WPI), University of Tokyo Institutes for Advanced Study, University of Tokyo, Kashiwa 277-8583, Japan}
\affiliation[e]{Istituto Nazionale di Fisica Nucleare, Sezione di Padova, Via Francesco Marzolo 8, 35131 Padova, Italy}

\abstract{
We point out that a QCD-like dark sector can be coupled to the Standard Model by gauging the topological Skyrme current, which measures the dark baryon number in the infrared,
to give a technically natural model for dark matter. This coupling allows for a semi-annihilation process $\chi \chi \rightarrow \chi X_\mu$, where $X_\mu$ is the gauge boson mediator and $\chi$ a dark pion field, which plays the dominant role in setting the dark matter relic abundance. The topological interaction is purely $p$-wave and so free from indirect detection constraints. We show that the dark matter pion mass needs to be in the range $10~{\rm MeV} \lesssim m_\chi \lesssim 1~{\rm TeV}$; towards the lighter end of this range, there can moreover be significant self-interactions. We discuss prospects for probing this scenario at collider experiments, ranging from the LHC to low-energy $e^+ e^-$ colliders, future Higgs factories, and beam-dump experiments.  
}

\begin{document}
\maketitle
~\newpage

\section{Introduction}

Understanding the nature of dark matter (DM), and how it interacts with Standard Model (SM) particles, are central challenges in fundamental physics~\cite{Cirelli:2024ssz}. The absence of any clear experimental clues coming from (in)direct detection or collider experiments motivates us to look beyond the most simple models for DM. 
And given the rich structure we observe in the visible sector, wherein the stable electron and proton are accompanied by a zoo of unstable particles, a compelling scenario is that DM comprises a few stable particles belonging to some similarly rich dark sector (DS). 
One appealing possibility is that of a confining QCD-like dark sector~\cite{Bai:2013xga}, say a dark $SU(N_c)$ gauge group acting on $N_f$ dark quarks $Q_i$ in the fundamental representation. For suitably chosen $N_c$ and $N_f$, we expect the infrared (IR) dynamics to be governed by a chiral symmetry breaking transition that delivers weakly-interacting pions $\chi^a$, some of which can be stable~\cite{Kribs:2016cew} and thus DM candidates.

Of course, once we go down this road we open up a Pandora's box of possible DS theories. It is pragmatic to focus on the different {\em portals} through which the DS can communicate with the SM, of the form $\mathcal{L}=\mathcal{O}_{\text{SM}}\mathcal{O}_{\text{DS}}$. Such a portal interaction (i) enables thermalisation of the DM with the SM, and so determines the relic abundance {\em e.g.} via freeze-out, and (ii) provides channels through which DM might be produced today in experiment. 
For example the vector portal~\cite{Holdom:1985ag}, by which a dark photon $X_\mu$ talks to the SM through kinetic mixing with the photon, is one of only three renormalisable portal interactions~\cite{Patt:2006fw,Falkowski:2009yz}. 

Given the hypothesis of a confining QCD-like sector, perhaps the most natural way to couple it to the SM is to gauge the dark baryon number current $j_{B,\text{UV}}^\mu = \frac{1}{N_c}\sum_i \overline{Q}_i\gamma^\mu Q_i$, which is an anomaly-free global symmetry of the DS, then couple to the SM via the vector portal, {\em viz.}
\begin{equation} \label{eq:intro-portalUV}
    {\cal L}_{\text{portal,UV}} = e_B X_\mu j_{B,\text{UV}}^\mu - \frac{\epsilon}{2} X_{\mu\nu}F^{\mu\nu}  
\end{equation}
where $X_{\mu\nu}=\partial_{[\mu}X_{\nu]}$ is the dark baryon field strength and
$F_{\mu\nu}=\partial_{[\mu}A_{\nu]}$ is the photon field strength.
Somewhat surprisingly, this simple possibility has been largely overlooked, and is the subject of this paper. We find its low-energy phenomenology has many striking features.

The key realisation is an old one: namely that, upon flowing to the IR, baryon number in a confining $SU(N_c)$ gauge theory can be identified with the topological winding number measuring the homotopy class of the pion field configuration~\cite{Skyrme:1961vq,Goldstone:1981kk,Balachandran:1982dw,Witten:1983tx}, that is an element in $\pi_3(SU(N_f)_L \times SU(N_f)_R/SU(N_f)_V) \cong {\mathbb Z}$. This is obtained by integrating the current
\begin{equation} \label{eq:intro-jB}
    j_{B,\text{IR}}^\mu = \frac{1}{24\pi^2}\epsilon^{\mu\nu\rho\sigma} 
    {\rm Tr}\left(U^{-1}\partial_\nu U\,U^{-1}\partial_\rho U\, U^{-1}\partial_\sigma U \right)\,,
\end{equation}
where $U(x)=e^{2i\chi^a(x) t^a/f_\pi}$ is the $SU(N_f)$-valued dark pion field, with $f_\chi$ being the corresponding decay constant.
The portal coupling in~\eqref{eq:intro-portalUV} therefore becomes a {\em topological interaction} in the IR, that is, loosely speaking, a term in the action that is obtained by directly integrating a differential form,\footnote{In the context of dark sector theories, topological interactions have previously been used (i) to realise the strongly interacting massive particle (SIMP) DM paradigm~\cite{Hochberg:2014dra} (by achieving $3\to2$ conversion within a sector of dark pions through the WZW term~\cite{Hochberg:2014kqa}), and (ii) to form a direct `topological portal' that converts QCD pions (and $\pi^0 \gamma$) to a pair of dark pions living on $SU(2)/SO(2)\cong S^2$~\cite{Davighi:2024zip,Davighi:2024zjp}. }
similar to the Wess--Zumino--Witten (WZW) term in QCD~\cite{Wess:1971yu,Witten:1983tw}.
Locally, {\em i.e.} for small field values, we can write the Lagrangian 
\begin{equation} \label{eq:intro-portalIR}
    {\cal L}_{\text{portal,IR}} = \frac{e_B}{12\pi^2 f_\chi^3} \epsilon^{\mu\nu\rho\sigma} f_{abc} X_\mu 
    \partial_\nu \chi^a \partial_\rho \chi^b \partial_\sigma \chi^c
    -\frac{\epsilon}{2}  X_{\mu\nu}F^{\mu\nu}  + O(\chi^4)\, ,
\end{equation}
where $f_{abc}$ are the structure constants for $SU(N_f)$.
Thus the dark pions $\chi^a$, which we expect to be stable and thus serve as our DM candidate, talk to the SM via a 4-point interaction involving three pions and one gauge boson.

We land on a DS theory in which dark pions talk to the vector $X$, which is itself thermalised with the SM through the vector portal, through a $2\to 1$ number-changing process. An interaction of this kind, $\chi\chi\to \chi X$, can realise the DM thermal freeze-out scenario known as `semi-annihilation'~\cite{DEramo:2010keq}. 
There is no need to impose {\em ad hoc} $\Z_3$ global symmetries to dictate this $2\to 1$ interaction structure; here it follows unavoidably from the fact that, given we seem to live in a three-dimensional space, a topological current corresponds to an invariant, closed differential 3-form.
Moreover, there is no contribution to the elastic channel $\chi\chi XX$ arising at tree-level, since the gauge field $X_\mu$ (and hence $A_\mu$) talks to the dark sector purely through the abelian gauge interaction whose structure is fixed to be linear in $X_\mu$. 
Operators inducing $\chi\chi XX$ transitions are expected to arise at 1-loop, originating from a dark quark box diagram in the UV theory which we estimate below, and so are subleading with respect to the $\chi\chi\chi X$ channel.
Our simple setup, namely dark QCD with gauged baryon number, therefore provides a natural UV completion for semi-annihilation.\footnote{Another natural UV completion for semi-annihilation exploits the structure of the Yang--Mills interaction, namely that there are cubic interactions between gauge bosons. Including also a charged scalar that talks to the SM via the scalar portal gives rise to the `hidden vector dark matter' model~\cite{Hambye:2008bq}. A scenario where semi-annihilation of DM to anti-DM explains the baryon asymmetry of the universe was discussed in \cite{Ghosh:2020lma}. }

In addition to being able to fit the observed DM relic abundance through thermal freezeout,\footnote{The same type of vertex can also lead to an exponential growth in the dark matter abundance through freeze-in \cite{Kramer:2020sbb}, which allows for a very heavy dark matter.} 
the composite nature of dark matter allows for a significant
velocity-dependent self-interaction amongst the dark pions, that offers a possible explanation~\cite{Kamada:2016euw,Rocha:2012jg,Peter:2012jh} of various small scale structure puzzles regarding galaxy formation~\cite{Flores:1994gz,Navarro:1995iw,Moore:1999gc,Moore:1999nt,Klypin:1999uc,Boylan-Kolchin:2011qkt,Boylan-Kolchin:2011lmk,Oman:2015xda} -- sometimes known as the `core-cusp problem' or the `diversity problem.'

Another striking feature of this model is that, because the interaction is topological and thus equipped with three derivatives acting on the dark pion fields~\eqref{eq:intro-portalIR}, the semi-annihilation process is necessarily $p$-wave. Without any tuning, this framework therefore evades otherwise strong constraints that would come from cosmic microwave background (CMB) observations \cite{Kawasaki:2021etm}, and indirect detection constraints due to gamma-ray emission from the galactic center \cite{Abazajian:2020tww}. This allows the DM mass to be light, into the sub-GeV range.

In this paper we explore the phenomenology of this elegant DS realisation of freeze-out by semi-annihilation, through the topological interaction~\eqref{eq:intro-portalIR} obtained by gauging baryon number.
We numerically solve the Boltzmann equations, including both the tree-level $\chi\chi\leftrightarrow\chi X$ semi-annihilation process and our estimate of the 1-loop induced $\chi\chi \leftrightarrow XX$, to obtain the region of parameter space for which the dark pions compose the whole DM relic abundance. This depends on the dark QCD parameters $m_\chi$, $f_\chi$, as well as the gauge coupling $e_B$ that activates the portal to the SM. Interestingly, we find viable parameter space where, in addition to fitting relic abundance through semi-annihilation, the following are all satisfied: (i) $m_\chi < 4\pi f_\chi$ for EFT validity; (ii) $m_\chi$ and $m_X$ are not so light as to modify $N_{\text{eff}}$ at the time of Big Bang nucleosynthesis (BBN); and (iii) the self-interactions are within the limits from astrophysical observations. At the edge of the range, these  self-interactions can be large enough to address the issues with small-scale structure. 

We conclude by surveying the prospects for probing this scenario in various collider experiments, for instance through classic mono-photon and mono-jet searches but also through more exotic dark shower signatures. These collider probes take on a greater importance given that direct and indirect detection probes are nullified by the $p$-wave nature of the interaction.

The rest of the paper is structured as follows. In \S \ref{sec:model} we set out the main features of the topological freeze-out model. In \S \ref{sec:DS_int} we compute the thermally-averaged cross-section for the semi-annihilation process, as induced at tree-level by the topological interaction, and also estimate the loop-induced annihilation channel. With these ingredients, we turn to solving the Boltzmann equations relevant for this dark matter model in \S \ref{sec:boltzmann}, providing both numerical and semi-analytical solutions (which are in good agreement). We discuss collider phenomenology in \S \ref{sec:pheno}, before concluding in \S \ref{sec:conclusion}. 

%%%%%%%%%%%%%%%%%%%%%%%%%%%%%%
%%%%%%%%%%%%%%%%%%%%%%%%%%%%%%
\section{The Model}
\label{sec:model}
%%%%%%%%%%%%%%%%%%%%%%%%%%%%%%
%%%%%%%%%%%%%%%%%%%%%%%%%%%%%%

Our model for the Dark Sector (DS) is simple: we posit an $SU(N_c)$ gauge theory with $N_f$ flavours of Dirac fermion $Q_i$ charged in the fundamental representation, with $U(1)_B$, the dark baryon number, also gauged. The Lagrangian reads
\begin{align} \label{eq:bigL}
    {\cal L}_{\rm UV} &= {\cal L}_{\mathrm{SM}} -\frac{1}{2}{\rm tr} G_{\mu\nu}G^{\mu\nu} -\frac{1}{4}X_{\mu\nu}X^{\mu\nu} - m_X^2 X_\mu X^\mu
    + \sum_i\bar{Q}_i (i{\not\!\!D} - m_Q) Q_i -\frac{\epsilon}{2 \cos\theta_w}  X_{\mu\nu}B^{\mu\nu},
\end{align}
where $D_\mu = \partial_\mu - i g G_\mu - i e_B X_\mu$, with $G_\mu$ the dark $SU(N_c)$ gauge field and $X_\mu$ that of $U(1)_B$; here we write the kinetic mixing in a way that is valid above the electroweak scale, with $B_{\mu\nu}$ being the hypercharge field strength and $\theta_w$ the weak mixing angle.
We assume there are at least two flavours $N_f \geq 2$ of dark quark $Q_i$, with a degenerate mass $m_Q$ that is smaller than the dynamical scale of $SU(N_c)$, $m_Q
< \Lambda_{N_c}$. Then below the strong scale the dark quarks are bound into pseudoscalar mesons $\chi^a$, which we generically refer to as `(dark) pions' and which serve as the dark matter in this model. For simplicity, we assume $N_f=2$ for much of the discussion below.
We include also a mass term for the dark photon, which could arise via the Higgs mechanism after spontaneously breaking $U(1)_B$ at a scale $\sim m_X/e_B$, or via the St\"uckelberg trick.

It might be surprising that the $U(1)_B$ gauge symmetry has anything to do with the dark pions because they do not carry baryon number. Yet, as we described in the Introduction, it was pointed out in \cite{Goldstone:1981kk} and later developed in \cite{Balachandran:1982dw,Witten:1983tx} that the low-energy baryon number current is given by
\begin{align} \label{eq:jB}
    j^\mu_B &= \frac{1}{24\pi^2} \epsilon^{\mu\nu\rho\sigma} 
    {\rm tr}(U^{-1}\partial_\nu U)(U^{-1}\partial_\rho U)(U^{-1}\partial_\sigma U)\, ,
\end{align}
as per \eqref{eq:intro-jB}.
Thus, adding the gauge interaction $e_B X_\mu j_B^\mu$
to the standard chiral Lagrangian
\begin{align}
    {\cal L}_\chi &= \frac{f_\chi^2}{4} {\rm tr} \partial^\mu U^\dagger \partial_\mu U
    + \mu^3 m_Q ({\rm tr} U + c.c.)\, , \qquad U(x)=e^{2i\chi^a(x) t^a/f_\chi}\, ,
\end{align}
where $\mu^3 = \langle \bar{Q} Q\rangle$,
and expanding it to the lowest order in $\chi^a$, we obtain
\begin{align}
    {\cal L}_\chi &= \frac{1}{2} (\partial^\mu \chi^a \partial_\mu \chi^a)
    - \frac{1}{2} m_\chi^2 \chi^a \chi^a
    + \frac{e_B}{12\pi^2 f_\chi^3} \epsilon^{\mu\nu\rho\sigma} f^{abc} X_\mu 
    \partial_\nu \chi^a \partial_\rho \chi^b \partial_\sigma \chi^c
    + O(\chi^4).
    \label{eq:Lag_chi}
\end{align}
The last term is a topological $\chi\chi\chi X$ vertex and causes semi-annihilation. Because we restrict to the case $N_f=2$, the model does not feature the more familiar WZW topological term involving five dark pion fields. It would be interesting to relax this assumption in future work.

\subsection{'t Hooft Naturalness and Dark Pion Stability}

The model is technically natural in the sense of 't Hooft~\cite{tHooft:1979rat}. There are no elementary scalars to cause a hierarchy problem. The assumption $m_Q < \Lambda_{N_c}$ is natural given that the limit $m_Q\rightarrow 0$ results in an enhancement of the symmetry, namely full chiral symmetry. The degeneracy among quarks can be ensured by a global flavour symmetry. Note that this degeneracy is important in order to ensure stability of the dark matter pions. For instance, the neutral dark pion $\chi^3 = (Q_1\bar{Q}_1-Q_2\bar{Q}_2)/\sqrt{2}$ is stable because the triangle diagrams involving up-quark and down-quark cancel precisely as long as $m_{Q_1} = m_{Q_2}$. If they are not degenerate, the difference between them lets $\chi^3$ decay into $XX$ if $m_X < m_\chi/2$, or off-shell into $X e^+ e^-$ {\em etc} if $m_\chi/2 < m_X < m_\chi$. 
If we are concerned that quantum gravity effects do not allow for a global flavour symmetry (see {\it e.g.} \cite{Reece:2023czb} and references therein), it can be an $SU(N_f=2)$ gauge symmetry with small coupling $g \lesssim (8\pi m_\chi/M_{\rm Pl})^{1/4}\sim 10^{-5} (m_\chi/{\rm 100~MeV})^{1/4}$ to avoid the gauge bosons to be populated in the thermal bath, or it can be a discrete gauge symmetry such as $S_3$ where $(u,d)$ is an irreducible doublet representation (see, {\it e.g.}\/, \cite{Hall:1995es}).

\subsection{Generalised Symmetry Matching}

Before we proceed to elucidate the phenomenological consequences of this DS portal, we pause to highlight more formal aspects of its construction. More phenomenologically-minded readers may wish to skip ahead to \S\ref{sec:DS_int}. In this Subsection, we briefly entertain the more general possibility of an arbitrary number $N_f$ of dark quark flavours.

Topological interactions in chiral Lagrangians play a key role when it comes to understanding the (possibly anomalous) symmetry structure of the theory in the infrared phase.
By integrating by parts, the local Lagrangian expression for the $\chi\chi\chi X$ term in~\eqref{eq:Lag_chi} can be written in a manifestly gauge-invariant way, that is in terms of the field strength $X_{\mu\nu}$ rather than $X_\mu$, as $\mathcal{L}\sim \epsilon^{\mu\nu\rho\sigma}\chi^a \partial_\mu \chi^b \partial_\nu \chi^c X_{\rho\sigma}$; but this Lagrangian is no longer manifestly invariant under the global flavour symmetry $SU(N_f)_L \times SU(N_f)_R$ which shifts the pion fields. This mirrors the story for the more familiar WZW term that also appears in our dark sector; in both cases, this quasi-invariance is strictly governed by the anomaly/symmetry structure of the underlying UV theory, which in turn governs why both topological interactions have quantised coefficients (in appropriate units). 

To understand how this works for the less-familiar portal interaction, first note that both topological actions can be written in a manifestly invariant form at the expense of locality, {\em i.e.} by passing to one higher dimension and writing the action as an integral of a differential form over an open 5-manifold $Y$ whose boundary is the 4-dimensional spacetime $\Sigma=\partial Y$ \`a la Witten~\cite{Witten:1983tw}. Using the language of differential forms, we have 
 \begin{equation} \label{eq:top}
    S_{\mathrm{top}}[\Sigma=\partial Y] = \int_Y \frac{-iN_c}{480\pi^3} \mathrm{Tr} (U^{-1} dU)^5 + \int_Y \frac{1}{24\pi^2} \mathrm{Tr} (U^{-1} dU)^3 \wedge X,  
 \end{equation}
where here $X:=\frac{1}{2}X_{\mu\nu}dx^\mu \wedge dx^{\nu}$ is the field strength 2-form for the $U(1)_B$ gauge field, and where we have temporarily absorbed the pre-factor of $e_B$ by redefining the gauge kinetic term in ~\eqref{eq:bigL}. The presence of a dynamical abelian gauge field bestows our DS with a 1-form global symmetry~\cite{Gaiotto:2014kfa} $H^{[1]}=U(1)$, whose current is simply $j^{(2)}=X$, which is a closed 2-form. The objects charged under this 1-form symmetry are 't Hooft lines for $X_\mu$. 
Furthermore, the second topological interaction in~\eqref{eq:top}, which is the focus of this work, intertwines this 1-form symmetry with the $G^{[0]}=SU(N_f)_L \times SU(N_f)_R$ flavour symmetry to form what is known as a {\em 2-group global symmetry}~\cite{Kapustin:2013uxa,Cordova:2018cvg,Cordova:2020tij}. The same structure was recently observed for a closely related theory in~\cite{Davighi:2024zjp}.

While a detailed explanation is somewhat tangential to this work, a quick way to see this is to turn on minimally-coupled background fields for both $G^{[0]}$ and $H^{[1]}$, denoted $A_{L,R}^{(1)}$ and $B^{(2)}$ respectively. Then observe that the gauged action $S \supset \int_Y \frac{1}{8\pi^2}\mathrm{Tr} \left[A_L dA_L + \frac{2}{3}A_L^3 - (L \leftrightarrow R)\right]\wedge X + i\int_\Sigma X \wedge B^{(2)}$ is invariant under the flavour background gauge transformations $A_{L,R} \to A_{L,R} + D_{A_{L,R}} \lambda_{L,R}^{(0)}$ iff the 2-form gauge field shifts as $B^{(2)} \to B^{(2)} + d\Lambda^{(1)}+\frac{1}{4\pi} \mathrm{Tr}(\lambda_L^{(0)}dA_L-\lambda_R^{(0)} d A_R)$. This results in a modification of the Ward identities for the ordinary flavour symmetry currents $j_{L,R}^a$, even with the background gauge fields switched off~\cite{Cordova:2018cvg,Davighi:2024zjp}. In the UV theory of dark QCD, this generalised symmetry structure arises as a consequence of a non-vanishing triangle diagram involving 2 flavour legs and 1 abelian gauge leg.

The upshot is that, while the ordinary WZW term matches a 't Hooft anomaly in the underlying theory (noting that our DS does not have the analogue of QED being gauged, so it is not in this case an ABJ anomaly), the second topological interaction, which is obtained by gauging the dark baryon number, matches a generalised 2-group symmetry, as in~\cite{Davighi:2024zjp}. 
Indeed, the need to match this generalised symmetry in the infrared can be seen as a rigorous justification for the baryon number current being identified with the topological winding number in the first place.
Both the anomaly coefficient and the 2-group coefficient are RG invariants of the theory; equivalently, the couplings of the two topological interactions are quantised. The EFT matching onto these coefficients is leading-order exact, and can be computed in the weakly-coupled UV phase -- unlike other non-topological interactions in the chiral Lagrangian. 

At this point, we fix $N_f=2$ again for the remainder of the paper -- meaning, in particular, that the first term on the RHS of Eq.~\eqref{eq:top} is not present in our setup.

% 

%%%%%%%%%%%%%%%%%%%%%%%%%%%%%%
%%%%%%%%%%%%%%%%%%%%%%%%%%%%%%
\section{Dark Sector Processes}
\label{sec:DS_int}

In this Section, we discuss important processes in the dark sector that will govern the evolution of both dark and visible species in the early Universe. These are (i) semi-annihilation $\chi\chi\rightarrow \chi X$ (\S\ref{sec:semi_annihilation}), (ii) annihilation $\chi \chi \rightarrow XX$ (\S\ref{sec:annihilation}), (iii) thermalisation $X \chi \rightarrow X \chi$ (\S\ref{sec:kin_eq}), and (iv) self-heating (\S\ref{sec:self-heating}). In order to discuss (ii) and (iii), we need to estimate the size of the effective $\chi\chi X X$ interaction which we do in \S\ref{sec:chiPT}.

\subsection{Semi-Annihilation Channel}
\label{sec:semi_annihilation}
%%%%%%%%%%%%%%%%%%%%%%%%%%%%%%
%%%%%%%%%%%%%%%%%%%%%%%%%%%%%%

The chiral Lagrangian with a gauged $U(1)_B$ symmetry, given in Eq.~\eqref{eq:Lag_chi}, gives rise to the semi-annihilation process $\chi \chi \rightarrow \chi X$ thanks to the topological interaction in~\eqref{eq:Lag_chi}. Here we analyze this process quantitatively, and identify the region of parameter space in which the dark pions can account for the observed dark matter abundance.

The key quantity to compute is the thermally averaged cross-section, $\langle\sigma v\rangle_{\chi\chi\to \chi X}$, which determines the effective rate of the semi-annihilation process entering the Boltzmann equation for dark pions and, consequently, governs their relic abundance. Our goal is to determine whether there exists a region of parameter space in which the dark pions can constitute all the DM as a thermal relic, undergoing freeze-out via semi-annihilations with the dark photon.

Starting from the Lagrangian~\eqref{eq:Lag_chi},
the contact interaction involving three dark pions and one dark photon gives rise to the following Feynman rule
\begin{equation}
    \begin{tikzpicture}
    [baseline=-3.0ex]
  \begin{feynman}
    \vertex (a);
    \vertex [above left=of a] (i1) {$\chi^a$};
    \vertex [below left=of a] (i2) {$
    \chi^b$};
    \vertex [above right=of a] (j1) {$X_\mu$};
    \vertex [below right=of a] (j2) {$\chi^c$};
    \diagram* {
      (i1) -- [dashed] (a), 
      (i2)--[dashed] (a),
      (a) -- [boson] (j1),
      (a) -- [dashed] (j2),
    };
    \node [draw=black, circle, fill=black, inner sep=0.7mm] at (a) {};
  \end{feynman}
\end{tikzpicture}
\sim -\frac{e_B}{2\pi^2 f_\chi^3} f^{abc} \epsilon^{\mu\nu\rho\sigma} p_{1\nu}\, p_{2\rho}\, p_{3\sigma}\,,
\label{eq:semiannihilation}
\end{equation}
where $p_{1,2,3}$ denote the (all incoming) momenta of the pions. This can be used to construct the corresponding matrix element, and thence the cross-section upon performing the final-state phase-space integration
\begin{equation}
    \sigma_{\chi\chi\to\chi X} = \frac{N_f}{N_f^2-1}\frac{e_B^2}{1536\pi^5} \frac{s^2}{f_\chi^6} \left(1-\frac{m_\chi^2}{s}\right)^3 \left(1-\frac{4m_\chi^2}{s}\right)^{1/2}\,,
    \label{eq:semi_xsec}
\end{equation}
where $s$ is the center-of-mass energy of the $\chi\chi\to\chi X$ process.
The details of this calculation are presented in App.~\ref{app:semi_annihilation}.
Following Refs.~\cite{Gondolo:1990dk,Edsjo:1997bg}, we compute the thermally averaged cross section times the M{\o}ller velocity, $v=\sqrt{|\mathbf{v}_1-\mathbf{v}_2|^2 - |\mathbf{v}_1\times\mathbf{v}_2|^2}$, as the following integral
\begin{equation}
		\langle \sigma v
		\rangle_{\chi\chi\to\chi X} = \frac{\int_{4m_\chi^2}^{\infty}\sigma_{\chi\chi\to\chi X}\sqrt{s}(s-4m_\chi^2)\,{\rm{K}}_1(\sqrt{s}/T)\,{\rm{d}}s}{8m_\chi^4 T\,{\rm K}_2^2(m_\chi/T)}\,,
		\label{eq:thavg_xsec}
\end{equation}
with ${\rm K}_i$ being the modified Bessel functions of the $i$-th order. 
The result can be expressed in terms of Meijer-G functions~\cite{ADAMCHIK1995283}, and we give such expressions in App.~\ref{app:semi_annihilation} for completeness. 

Because the dark pions are nearly non-relativistic at the time of freeze-out, it is useful to Taylor expand the thermally averaged cross-section $\langle \sigma v \rangle_{\chi\chi\to\chi X}$ in inverse powers of $x := m_\chi/T$. 
We obtain
\begin{equation}
    \langle \sigma v
		\rangle_{\chi\chi\to\chi X} = \frac{N_f}{N_f^2-1} \frac{e_B^2}{384 \pi^{5}}\frac{m_\chi^4}{f_\chi^6} \left(\frac{81}{16} \,x^{-1} + \frac{891}{32}\,x^{-2} + \mathcal{O}(x^{-3})\right)\,.
        \label{eq:semiann_expanded}
\end{equation}
We will use this expression as input to the Boltzmann equation in \S\ref{sec:boltzmann}. There we will moreover show that the leading order term, proportional to $x^{-1}$ in Eq.~\eqref{eq:semiann_expanded}, captures the dominant contribution to the semi-annihilation rate, with deviations at the level of $\mathcal{O}(10\%)$ compared to the full numerical evaluation of Eq.~\eqref{eq:semi_thermal_avg_xsec} (from the Appendix).

The fact that the leading term in this expansion starts at order $x^{-1}$ means that the semi-annihilation process at threshold is purely $p$-wave. This can be understood straightforwardly in the case where the final state dark photon is transverse, with helicity $\pm 1$, simply because of angular momentum conservation: 
namely, there is one unit of net angular momentum flowing along the axis of the final state $\chi X$ system, and hence the total angular momentum must be at least one unit, while the initial state $\chi\chi$ system has angular momentum only due to its orbital angular momentum $L$. On the other hand, it is also easy to see that a longitudinal $X_\mu$ is not produced. If the semi-annihilation had been $s$-wave, it would be subject to strong constraints from CMB data \cite{Kawasaki:2021etm} and Fermi-LAT measurements of gamma ray emission coming from the galactic center, halos, or dwarf galaxies~\cite{Abazajian:2020tww}: our model for semi-annihilation is free from such constraints.

\subsection{Annihilation Channel}
\label{sec:annihilation}

In addition to the semi-annihilation process — which plays a central role in determining the dark pion abundance and arises from the topological interaction in Eq.~\eqref{eq:Lag_chi} — there are other relevant processes induced by additional operators in the chiral Lagrangian. In particular, it has been assumed in the literature that dark matter remains in kinetic equilibrium with the Standard Model during freeze-out via self-annihilation processes~\cite{DEramo:2010keq}. In our model, kinetic equilibrium can be maintained through scattering processes such as $X \chi \rightarrow X \chi$. The crossed diagram also contributes to the annihilation process $\chi \chi \rightarrow XX$, which may compete with the semi-annihilation channel. It is therefore essential to study the $\chi\chi XX$ coupling in general. 

\subsubsection{NDA Estimate}
\label{sec:chiPT}

Using na\"ive dimensional analysis, we estimate the size of the operators responsible for the $\chi\chi XX$ interaction as follows
\begin{align}
    \mathcal{L}_{\chi{\rm PT}} \supset \left(\frac{e_B}{16\pi^2 f_\chi}\right)^2 \left(\lambda_1 \left(\partial_\alpha U^\dagger\right) \left(\partial^\alpha U\right) X_{\mu\nu}\,X^{\mu\nu}+\lambda_2 \left(\partial_\alpha U^\dagger\right) \left(\partial^\nu U\right) X_{\mu\nu}\,X^{\mu\alpha}\right)\,,
    \label{eq:L_xhiPT}
\end{align}
with $\lambda_{1,2} \sim O(1)$. We next present estimates of these loop-induced couplings from both the IR and UV sides, and show they agree with this NDA expectation. We then discuss their consequences regarding the annihilation and thermalisation processes.

\subsubsection{IR Estimate}

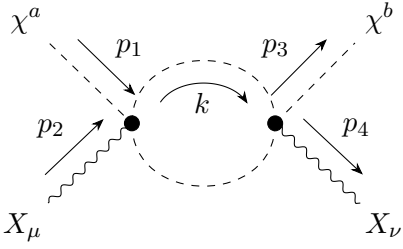
\begin{figure}[t]
\centering
\begin{tikzpicture}[baseline=(b.base)]
  \begin{feynman}
    \vertex (a);
    \vertex [above left=of a] (i1) {$\chi^a$};
    \vertex [below left=of a] (i2) {$X_\mu$};
    \vertex [right=0.75in of a] (b);
    \vertex [above right=of b] (j1) {$\chi^b$};
    \vertex [below right=of b] (j2) {$X_\nu$};
   
    \diagram* {
(a) -- [dashed, half left,momentum'=$k$] (b),
(b) -- [dashed, half left] (a),
      (i1) -- [dashed,momentum=$p_1$] (a), 
      (i2)--[boson,momentum=$p_2$] (a),
      (b) -- [dashed,momentum=$p_3$] (j1),
      (b) -- [boson,momentum=$p_4$] (j2),
    };
    \node [draw=black, circle, fill=black, inner sep=0.7mm] at (a) {};
\node [draw=black, circle, fill=black, inner sep=0.7mm] at (b) {};
  \end{feynman}
\end{tikzpicture}
\caption{One-loop contribution to elastic channel $\chi^a \chi^a \leftrightarrow XX$, mediated through two insertions of the topological interaction. 
 \label{fig:1loop_IR}}
\end{figure}

We here estimate the loop-induced contribution within the low-energy EFT to the elastic channel
    $\chi^a X_\mu \leftrightarrow \chi^b  X_\nu$.
Inserting twice the topological operator as shown in Fig.~\ref{fig:1loop_IR}, using the $\mathfrak{su}(N_f)$ Lie algebra identity
    $f_{acd}f_{cdb} = -N_f\delta_{ab}$
to sum over the flavour indices of the dark pions in the loop,
and doing an integral over the loop momenta, we find the following one-loop amplitude 
\begin{align}
    i\mathcal{M}_{\chi^a X\to \chi^b  X} = 
    &\left(\frac{e_B}{2\pi^2 f_\chi^3}\right)^2 
    N_f\,\delta_{ab} \epsilon^{\mu\lambda\rho\sigma} \epsilon^{\nu\alpha\beta\gamma} p_{1\sigma}p_{2\mu} \varepsilon_{2\lambda}(p_2) p_{3\gamma} p_{4\nu} \varepsilon_{4\alpha}(p_4) \nonumber \\
    &\times \int \frac{d^4k}{(2\pi)^4} \frac{k_\rho k_\beta}{(m_\chi^2-k^2)(m_\chi^2-(k-p_1-p_2)^2)} \,,
\end{align}
where $k$ is the loop momentum, and $p_{1,3}$ ($p_{2,4}$) are the dark pion (photon) momenta.
By Lorentz invariance, the loop integral can be decomposed into two pieces
\begin{equation}
     \int \frac{d^4k}{(2\pi)^4} \frac{k_\rho k_\beta}{(m_\chi^2-k^2)(m_\chi^2-(k-p_1-p_2)^2)} \equiv  \frac{1}{4}g_{\rho\beta} I  + (p_1+p_2)_\rho (p_3+p_4)_\beta J        \, .
\end{equation}
The piece $J$ will not contribute to the matrix element when contracted with the epsilon tensors, due to antisymmetry and the appearance of momenta in the matrix element written above. 
The important loop integral to keep is therefore the scalar integral
\begin{equation}
    I(s,m_\chi):=\int \frac{d^4k}{(2\pi)^4} \frac{k^2}{(m_\chi^2-k^2)(m_\chi^2-(k-p_1-p_2)^2)}\, ,
\end{equation}
which is a function of the dark pion mass $m_\chi$ and the kinematic Mandelstam $s=(p_1+p_2)^2$.

The loop integral $I(s,m_\chi)$ has mass-dimension 2, reflecting a na\"ive quadratic divergence. 
We can obtain an order of magnitude estimate by cutting off the loop integral at $k=\Lambda~ \gg m_\chi, \sqrt{s}$ and keeping the leading part coming from the quadratic divergence, giving $I(s,m_\chi) \approx \Lambda^2 / (4\pi)^2$. The coefficient becomes
\begin{align}
    \frac{N_f}{4}\left(\frac{e_B}{2\pi^2 f_\chi^3}\right)^2 \frac{\Lambda^2}{(4\pi)^2}
    \approx N_f \frac{e_B^2 \Lambda^2}{(4\pi f_\chi)^6}\ . 
    \label{eq:IR}
\end{align}
The cut-off dependence of this IR estimate suggests the $\chi\chi XX$ coupling comes from the strong dynamics.

\subsubsection{UV Estimate}

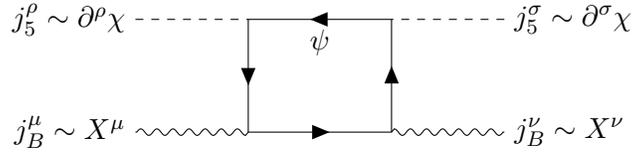
\begin{figure}[t]
\centering
\begin{tikzpicture}
  \begin{feynman}
    \vertex (a);
    \vertex [left=of a] (i1) {$j_5^\rho \sim \partial^\rho \chi$};
    \vertex [below=of a] (b);
    \vertex [left=of b] (i2) {$j_B^\mu \sim X^\mu$};
    \vertex [right=0.75in of a] (c);
    \vertex [right=of c] (j1) {$j_5^\sigma \sim \partial^\sigma \chi$};
    \vertex [right=0.75in of b] (d);
    \vertex [right=of d] (j2) {$j_B^\nu\sim X^\nu$};
   
    \diagram* {
(a) -- [fermion](b) -- [fermion](d) --[fermion] (c) --[fermion,edge label=$\psi$] (a),
      (i1) -- [dashed] (a), 
      (i2) -- [photon] (b),
      (j1) -- [dashed] (c),
      (j2) -- [photon] (d),
    };
  \end{feynman}
\end{tikzpicture}
\caption{A one-loop diagram contribution to the 4-point function $\langle j_5^\rho j_5^\sigma j_B^\mu j_B^\nu \rangle$ as estimated from the UV -- see main text. All permutations among the vertices need to be added. 
 \label{fig:1loop_UV}}
\end{figure}

Here we attempt to estimate the $\chi \chi XX$ coupling from the UV theory. The interpolating field for the pion is the axial current $f_\chi \partial^\mu \chi^a \approx j_5^{a\mu} = \bar{Q} \gamma^\mu \gamma_5 T^a Q$, and the $\chi \chi XX$ interaction comes from the box diagram of the quark loop $\langle j_5^\mu j_5^\nu j_B^\rho j_B^\sigma\rangle$ shown in Fig.~\ref{fig:1loop_UV}. Because of the gauge invariance, two powers of external momenta are extracted out of the integral, and it is quadratically IR divergent. Cutting it off at $\Lambda$ as the ``matching scale'' between the UV theory and the chiral Lagrangian, the coefficient is 
\begin{align}
    \frac{e_B^2}{N_c} \frac{1}{(4\pi)^2} \frac{1}{\Lambda^2} \frac{1}{f_\chi^2}
    = \frac{1}{N_c} \frac{e_B^2}{(4\pi f_\chi)^2 \Lambda^2}\ . \label{eq:UV}
\end{align}
Apart from model-dependent factors of $N_f,N_c$, Eqs.~\eqref{eq:IR} and \eqref{eq:UV} become the same if the theories are matched at $\Lambda = 4\pi f_\chi$, namely where the theory is strongly coupled. We consider this to be a good estimate.

%%%%%%%%%%%%%%%%%%%%%%%%%%%%%%
%%%%%%%%%%%%%%%%%%%%%%%%%%%%%%

\subsubsection{Annihilation Cross-Section}

We are now able to estimate the contribution of the annihilation channel to the evolution of the dark pion abundance using the effective operators in Eq.~\eqref{eq:L_xhiPT}, parametrising the unknown aspects of the strong dynamics through the corresponding Wilson coefficients $\lambda_1$ and $\lambda_2$, which we assume to be of order unity. 

As in the case of semi-annihilation, we defer the computational details to App.~\ref{app:annihilation}, and provide here the expression for the $\chi\chi \to XX$ cross-section and its corresponding thermal average. In the limit of the dark photon being massless, we find
\begin{align}
    \sigma_{\chi\chi\to XX} &= \frac{1}{N_f^2-1}\left(\frac{e_B}{16\pi^2 f_\chi^2}\right)^4\frac{1}{8\pi}\, s\sqrt{\frac{s}{s-4m_\chi^2}}\\ 
    &\times\left[(s-2m_\chi^2)^2\left({\rm Re}(\lambda_1\,\lambda_2^*)+ 2 |\lambda_1|^2\right)
    +\frac{1}{120} (92 m_\chi^4 - 76 m_\chi^2 s + 17 s^2) |\lambda_2|^2\right]\,.\nonumber
\end{align}
The full expression for the thermally averaged cross-section (still in the massless dark photon limit) is then given in Eq.~\eqref{eq:ann_thermal_avg_xsec}; as before, it is most instructive expand this result in powers of $x = m_\chi / T$, to obtain
  \begin{align}
        \langle \sigma v\rangle_{\chi\chi\to XX} &\approx \frac{1}{N_f^2-1}\left(\frac{e_B}{16\pi^2 f_\chi^2}\right)^4\frac{m_\chi^6}{2\pi} |4\lambda_1+\lambda_2|^2 \left(1+\frac{15}{8}x^{-1}+\mathcal{O}(x^{-2})\right)\,.
        \label{eq:sigmav_annihilation_nr}
    \end{align}
We observe that the annihilation process now contains a velocity-independent leading order piece, unlike what we found for semi-annihilation, meaning it is in principle subject to constraints coming from indirect detection and CMB data. However, in contrast to the tree-induced semi-annihilation process, the annihilation channels in this model are loop-suppressed and become relevant only in the regime where $m_\chi > 4\pi f_\chi$. We can assess this numerically. For instance, considering the CMB constraints, the annihilation cross section satisfies $\langle\sigma v\rangle_{\chi\chi\to XX} \lesssim 3\times 10^{-26}\,\text{cm}^3/\text{s}$, for $m_\chi>1$ GeV and $e_B<10^{-2}$, which remains below the sensitivity of the Planck satellite even when $m_\chi = 4\pi f_\chi$~\cite{Planck:2018vyg}. 

Generally, the relative importance of annihilations compared to semi-annihilations can be quantified via the ratio
\begin{equation}
    \frac{\langle\sigma v\rangle_{\chi\chi\to XX}}{\langle\sigma v\rangle_{\chi\chi\to \chi X}} \simeq \frac{|4\lambda_1+\lambda_2|^2\,e_B^2}{64\pi^4 N_f} \frac{m_\chi^2}{f_\chi^2}\,x\,,
\end{equation}
from which we see that annihilations can become comparable to semi-annihilations in determining the relic abundance when $m_\chi > 4\pi f_\chi$ assuming benchmark values $\lambda_{1,2}\sim\mathcal{O}(1)$, $e_B \sim 0.1$, and $x=x_f\sim\mathcal{O}(20)$ at freeze-out. Consequently, in the parameter space where dark pions can constitute viable dark matter candidates — namely $m_\chi < 4\pi f_\chi$, as discussed in the following section — the annihilation channel is expected to provide only a subdominant contribution. This expectation is explicitly confirmed in the numerical analysis of \S\ref{sec:putting_together}.

%%%%%%%%%%%%%%%%%%%%%%%%%%%%%%
%%%%%%%%%%%%%%%%%%%%%%%%%%%%%%
\subsection{Kinetic Equilibrium}
\label{sec:kin_eq}
%%%%%%%%%%%%%%%%%%%%%%%%%%%%%%
%%%%%%%%%%%%%%%%%%%%%%%%%%%%%%

The elastic scattering process $X \chi\to X\chi$ thermalises the dark pions with the Standard Model bath since the dark photon $X_\mu$ kinetically mixes with the ordinary photon. Throughout this paper, we assume that $X_\mu$ maintains thermal equilibrium with the Standard Model. It requires only a modest lower limit on the kinetic mixing 
\begin{equation} \label{eq:epsilon}
    \epsilon \gtrsim \alpha^{-1} \left(\frac{T}{M_{\rm Pl}}\right)^{1/2} \simeq 4\times 10^{-8} 
    \left(\frac{T}{\rm GeV}\right)^{1/2}
    \,,    
\end{equation}
where $\alpha$ is the electromagnetic fine-structure constant.

To identify the region of the $(m_\chi, f_\chi)$ parameter space where dark pion thermalisation is effective, we compare the interaction rate of the $X \chi\to X\chi$ process with the Hubble expansion rate. The thermalisation condition requires
\begin{equation}
    \Gamma_{X\chi\to X\chi} = n_\chi \langle \sigma v\rangle_{X\chi\to X\chi}\gtrsim H\,,
    \label{eq:Gamma_therm}
\end{equation}
where $n_\chi$ is the dark pion number density and $H$ is the Hubble parameter.

The cross-section can be obtained using the matrix elements in Eqs.~\eqref{eq:O1} and~\eqref{eq:O2} and the convolution with the final state phase-space. In the limit of the massless dark photon $X_\mu$, we find
\begin{align}
    \sigma_{X\chi \to X\chi} &= \frac{1}{64\pi s}\left(\!\frac{e_B}{16\pi^2 f_\chi^2}\!\right)^{\!4}\!\left(\!1-\frac{m_\chi^2}{s}\!\right)^{\!4}\nonumber\\
    &\times \Bigg[|4\lambda_1+\lambda_2|^2 \left(\frac{1}{5}(s-m_\chi^2)^4-(s-m_\chi^2)^2 s\, m_\chi^2 + \frac{4}{3}s^2 m_\chi^4\right)
    +\frac{4}{3} |\lambda_2|^2 s^4 \Bigg]\,.
    \label{eq:sigma_therm}
\end{align}
We are interested in whether the kinetic equilibrium between the dark pions and the SM bath is kept long enough until the onset of the dark pion freeze-out, which occurs at temperatures $T \sim m_\chi/ x_f$ with $x_f \sim \mathcal{O}(10-30)$ depending on the values of the parameters (with larger $x_f$ achieved for larger dark pion masses). At these temperatures, we consider the dark pions to be non-relativistic, such that their relative velocity with dark photons is $v\sim 1$, and the center-of-mass energy can be approximated by
\begin{equation}
    s = (p_\chi+p_X)^2 = m_\chi^2 + 2 E_\chi E_X (1-\cos\theta) \stackrel{\int d\theta}{\simeq} m_\chi^2 + 2m_\chi T\,,
\end{equation}
where we used that $E_X\sim T$, $E_\chi \sim m_\chi + 3T/2$, and we integrated over the angle $\theta$ between the $\chi$ and $X$ momenta assuming the uniform distribution. 

Substituting this into Eq.~\eqref{eq:Gamma_therm} and using the assumed equilibrium distribution for the dark pions, their successful thermalisation at $T \sim m_\chi/ x$ requires
\begin{align}
    \frac{(|4\lambda_1+\lambda_2|^2+|\lambda_2|^2)\,e_B^4}{12288\sqrt{2}\pi^{21/2} x^{11/2}}\, e^{-x} \frac{m_\chi^9}{f_\chi^8} \gtrsim \frac{m_\chi^2}{x^2 M_{\rm Pl}}\,.
    \label{eq:therm_cond}
\end{align}
The region of the $(m_\chi,f_\chi)$ parameter space where kinetic equilibrium is not maintained through the elastic scattering channel is indicated by various shades of purple in Fig.~\ref{fig:semi-ann-full}. Each shade corresponds to the parameter space for which kinetic equilibrium persists only up to a given stage in the cosmological evolution, quantified by $x_f = m_\chi/T_{\rm FO}$, with $T_{\rm FO}$ denoting the freeze-out temperature. The figure illustrates that, in general, elastic scattering via $\chi X\to \chi X$ is insufficient to sustain kinetic equilibrium between $\chi$ and the SM bath right up to the freeze-out moment. As a result, the DM temperature need not track that of the SM bath, potentially leading to a non-trivial thermal history for the dark sector~\cite{Kamada:2017gfc} that we should account for, which we do in the following Subsection.

\subsection{Self-Heating}
\label{sec:self-heating}

Ref.~\cite{Kamada:2017gfc} studied the consequences of DM going out of thermal equilibrium before freeze-out via semi-annihilation is completed, which will happen if the elastic $\chi X \to \chi X$ scattering is not efficient enough. 
They showed that, if the semi-annihilation is $s$-wave, the $\chi \chi \to X \chi$ process starts to inject significant energy into the final state $\chi$-particle as $\chi$ becomes non-relativistic, specifically $E_{\chi, f} = m_\chi \left(\frac{5}{4} - \frac{m_X^2}{4m_\chi^2} \right) \equiv \gamma m_\chi$, leading to an increase in the DM temperature {\em viz.} 
${T_\chi}/{T_{\rm SM}} \approx \frac{2}{3} (\gamma-1) x_f$ that can be as large as a factor $5$ to $6$ enhancement.

However, the authors of Ref.~\cite{Kamada:2021muh} concluded that there is no significant self-heating when the semi-annihilation is in the $p$-wave, which is the case for our model. This is because the kinetic energy density of dark matter $\chi$ redshifts adaibatically as
\begin{align}
    \rho_K = \left\langle \frac{p^2}{2m_\chi} \right\rangle n_\chi \propto a^{-5},
\end{align}
while the injection due to the semi-annihilation is
\begin{align}
    \Delta \rho_K \approx \frac{1}{H} \langle \sigma v \rangle n_\chi^2
    \propto a^2 a^{-2L} a^{-6}
\end{align}
which is important for $L=0$ ($s$-wave) at later times but not for $L \geq 1$ ($p$-wave and beyond). 
At the same time, the final abundance of dark matter is not modified more than 10\% by going out of equilibrium (in either the $s$-wave or $p$-wave case). 

Therefore, we will not discuss the possibility of self-heating any further in our paper, and we are reassured that even in parameter space regions for which thermalisation is not maintained we do not expect a large correction to the DM relic abundance.

%%%%%%%%%%%%%%%%%%%%%%%%%%%%%%f
%%%%%%%%%%%%%%%%%%%%%%%%%%%%%%
\section{Boltzmann Equation}
\label{sec:boltzmann}
%%%%%%%%%%%%%%%%%%%%%%%%%%%%%%
%%%%%%%%%%%%%%%%%%%%%%%%%%%%%%

In this section, we solve the Boltzmann equation governing the evolution of the dark pion number density $n_\chi$, taking into account the processes considered in the previous Section, and demonstrate that dark pions can account for the observed dark matter abundance. We identify the combinations of model parameters that yield the correct relic density and delineate the viable region of parameter space. This region is bounded by several theoretical and observational constraints: the validity of the EFT, limits on $N_{\rm eff}$, bounds from dark pion self-interactions, and the requirement that kinetic equilibrium with the Standard Model is maintained long enough to ensure the robustness of our predictions.

In the presence of dark pion annihilations ($\chi\chi\to XX$) and semi-annihilations ($\chi\chi\to\chi X$), the Boltzmann equation for the dark pion number density $n_\chi$ reads
\begin{equation}
    \frac{dn_\chi}{dt}+3Hn_{\chi} = -\langle\sigma v\rangle_{\chi\chi\to XX}\left(n_\chi^2-n_\chi^{{\rm eq}\,2}\right) -\frac{1}{2}\langle\sigma v\rangle_{\chi\chi\to \chi X}\left(n_\chi^2-n_\chi n_\chi^{{\rm eq}}\right)\,,
    \label{eq:Boltzmann_n_pi}
\end{equation}
where $H$ is the Hubble parameter and $n_\chi^{{\rm eq}}$ is the equilibrium number density distribution. For convenience, it is more suitable to rewrite Eq.~\eqref{eq:Boltzmann_n_pi} as 
\begin{equation}
    \frac{dY_{\chi}}{dx} = -\sqrt{\frac{\pi g_*}{45}}\frac{M_{\rm Pl} m_{\chi}}{x^2}\left[\langle\sigma v\rangle_{\chi\chi\to XX}\left(Y_{\chi}^2-Y_{\chi}^{{\rm eq}\,2}\right)+\frac{1}{2}\langle\sigma v\rangle_{\chi\chi\to \chi X}\left(Y_{\chi}^2-Y_{\chi}Y_{\chi}^{{\rm eq}}\right)\right]\,,
    \label{eq:Boltzmann_yield}
\end{equation}
where we defined a comoving number density, or yield, $Y_\chi = n_\chi/s$, with $s$ being the entropy density of the relativistic degrees of freedom, and the evolution parameter is $x=m_\chi/T$.
Here $M_{\rm Pl}$ is the Planck mass, $g_*$ is the effective number of relativistic degrees of freedom, and the equilibrium yield $Y_\chi^{\rm eq}$ is a function of the evolution parameter $x$, given by
\begin{equation}
    Y_{\chi}^{\rm eq} = \frac{45}{4\pi^4 g_*} x^2 {\rm K}_2(x)\,,
    \label{eq:eq_yield}
\end{equation}
with ${\rm K}_2(x)$ being the modified Bessel function of the second order. The factor of one half in the semi-annihilation contribution stems from the symmetry factors that
account for identical particles in the initial and final states of the collision operator (see \emph{e.g.} Ref.~\cite{DEramo:2025xef}).

We solve Eq.~\eqref{eq:Boltzmann_yield} numerically by substituting the explicit expressions for the thermally averaged cross-sections $\langle\sigma v\rangle_{\chi\chi\to \chi X}$ and $\langle\sigma v\rangle_{\chi\chi\to XX}$ in Eqs.~\eqref{eq:semi_thermal_avg_xsec} and~\eqref{eq:ann_thermal_avg_xsec}, respectively. We find that there is indeed a viable parameter space that allows us to set the correct relic abundance for dark pions
\begin{equation}
		\Omega_\chi h^2 \approx \frac{1\, m_\chi\, Y_{\infty}}{3.6\cdot 10^{-9} \,{\rm GeV}} \approx 0.12\,,
  \label{eq:correct_relic}
\end{equation}
with $Y_{\infty}$ defined as a solution of Eq.~\eqref{eq:Boltzmann_yield} in the limit $x\to \infty$. 

\begin{figure}[t]
    \centering
    \includegraphics[width=0.8\linewidth]{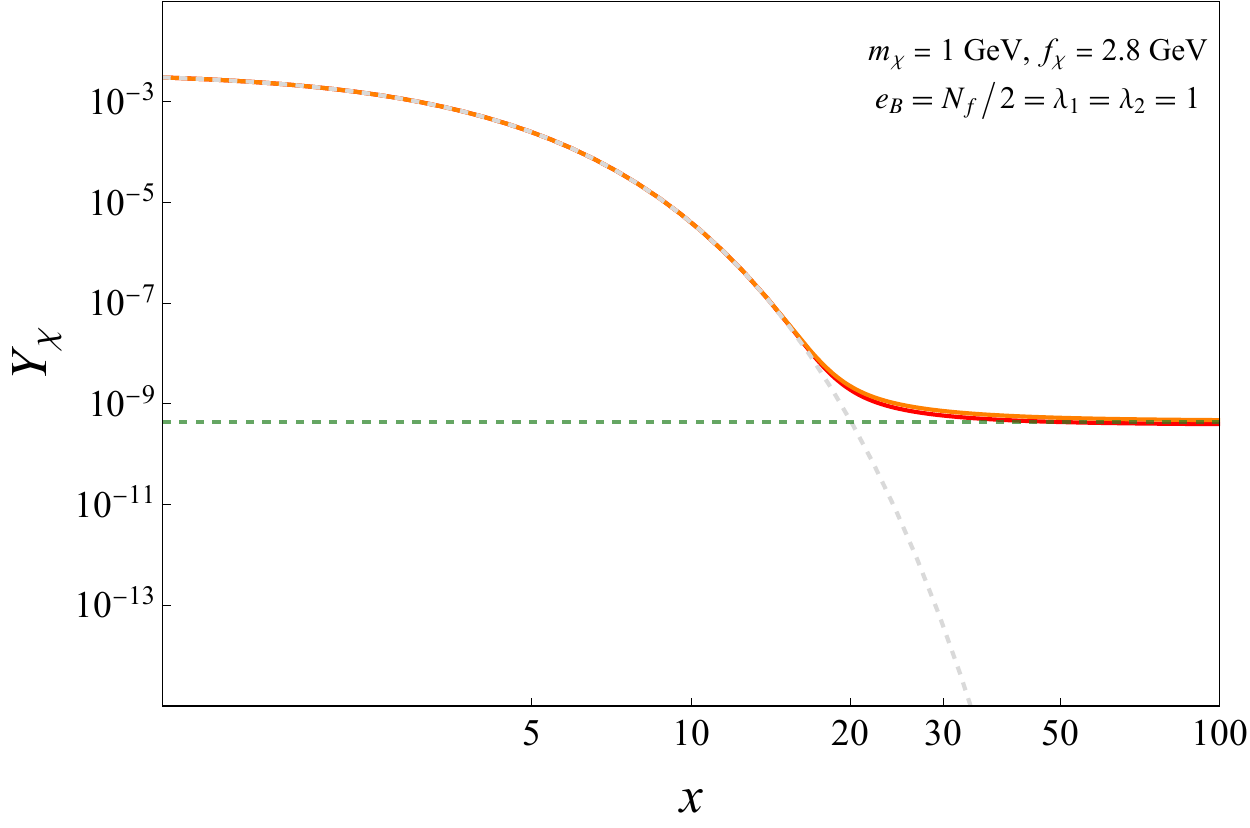}
    \includegraphics[width=0.8\linewidth]{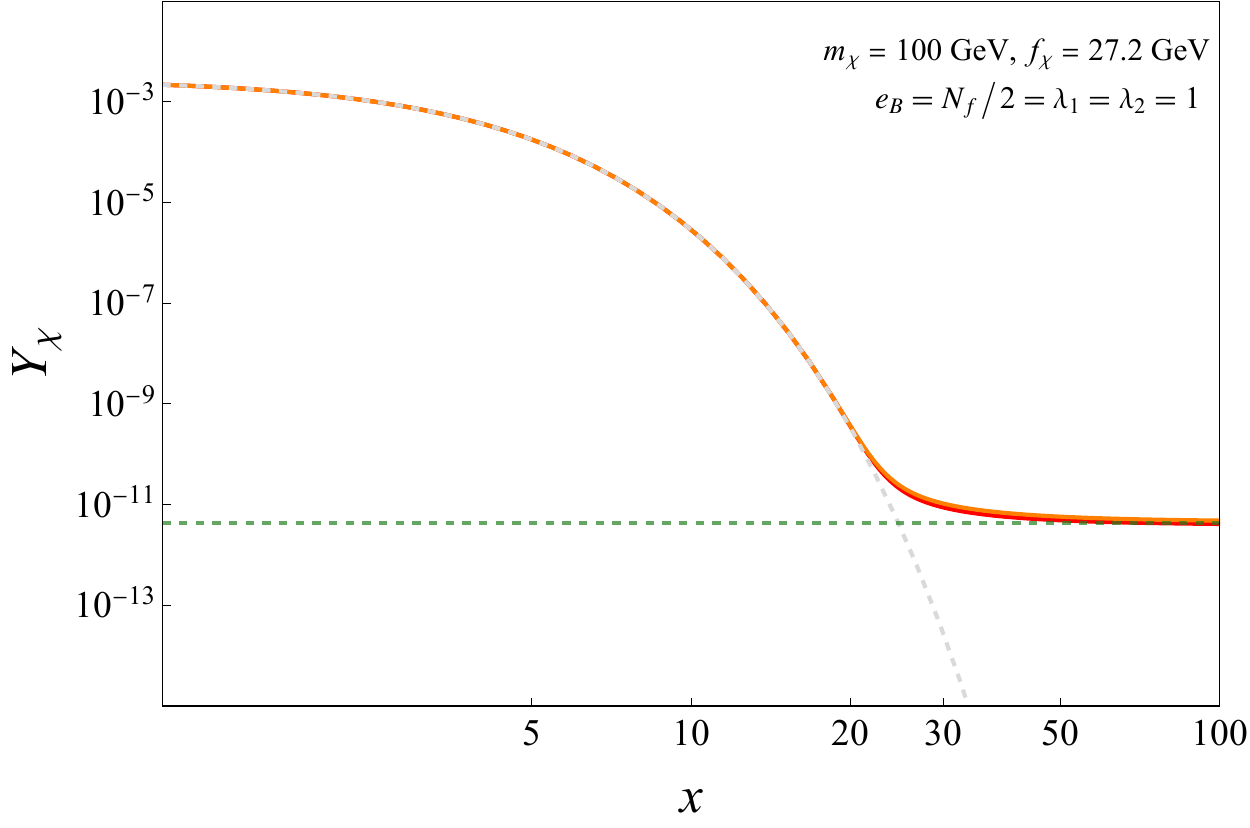}
    \caption{ 
    The dark pion yield $Y_\chi$ as a function of $x=m_\chi/T$ for two benchmark parameter sets. The solid red (orange) line represents the numerical solution of the Boltzmann equation~\eqref{eq:Boltzmann_yield} using the full (leading-order) $x$-dependence of the thermally averaged cross sections. The observed relic abundance is achieved for the $Y_\chi$ value denoted by the dashed green line, while the grey dashed line shows the equilibrium distribution $Y_{\chi}^{\rm eq}(x)$.}
    \label{fig:Boltzmann_evo}
\end{figure}

We show the evolution of $Y_\chi$ for two benchmark values of the model parameters in Fig.~\ref{fig:Boltzmann_evo}, namely: 
\begin{align}
    \text{Benchmark 1:} \quad & m_\chi = 1 \text{~GeV}, \,\,\,\,\,\,\,\, f_\chi = 2.8 \text{~GeV} , \\
    \text{Benchmark 2:} \quad & m_\chi = 100 \text{~GeV}, \,\, f_\chi = 27.2 \text{~GeV} ,
\end{align}
and taking $e_B = N_f/2=\lambda_1=\lambda_2=1$ for both benchmarks.
The solid red line represents the numerical solution of the Boltzmann equation using the full $x$-dependence of the thermally averaged cross sections. In contrast, the solid orange line corresponds to the solution obtained by retaining only the leading-order term in the velocity expansion of Eqs.~\eqref{eq:semiann_expanded} and~\eqref{eq:sigmav_annihilation_nr}. The resulting difference in the final relic abundance is at the level of 10\% - 15\%, consistent with the expectation that dark pions are nearly non-relativistic at freeze-out.

In the following Subsections we compare the contributions of semi-annihilation and annihilation processes in determining the final dark pion abundance. We show that, within the relevant region of parameter space, annihilation processes are subdominant and can be safely neglected. This simplification enables us to obtain a more transparent analytical understanding of the parameter dependencies, and clarifies the conditions under which dark pions constitute a viable dark matter candidate.

\subsection{Semi-Annihilation Contribution}

Let us start by analysing the Boltzmann equation keeping only the contribution from the semi-annihilation process, {\em i.e.} by setting $\langle \sigma v\rangle_{\chi\chi\to XX}= 0$ in the Boltzmann equation~\eqref{eq:Boltzmann_yield}. 
To better understand how various parameters affect the final abundance, we can use the semi-analytic solution for $\Omega_\chi h^2$ derived in~\cite{DEramo:2010keq}. For the case of semi-annihilations, the freeze-out value $x_f$ can be found by solving the equation
\begin{equation}
    x_f = \log\left[0.038 c(c+1) \frac{m_\chi M_{\rm Pl}}{\sqrt{g_* x_f}} \frac{1}{2}\langle \sigma v\rangle_{\chi\chi \to \chi X}\right]\,,
    \label{eq:xf_semi-annihilation}
\end{equation}
where $c \in \mathbb{R}$ is a constant that encodes the information on the offset of the freeze-out; it is defined numerically through 
\begin{equation} \label{eq:c}
    Y_\chi(x_f)-Y_\chi^{\rm eq}(x_f)=c\, Y_\chi^{\rm eq}(x_f)\, .
\end{equation}
The value of $x_f$ exhibits only a logarithmic dependence on $\langle \sigma v\rangle_{\chi\chi \to \chi X}$, and, for values of the model parameters yielding the correct relic abundance, it typically lies in the range $x_f\in[15,30]$. Its precise value can be determined numerically and subsequently used to compute the present-day dark pion energy density
\begin{equation}
    \Omega_\chi h^2 = 2\times \frac{1.07\times 10^{9}\, {\rm GeV^{-1}}}{\sqrt{g_*} M_{\rm Pl} J(x_f)}\,,
    \label{eq:omega_pi_h2}
\end{equation}
where
\begin{equation}
    J(x_f) = \int_{x_f}^\infty \frac{\langle \sigma v \rangle_{\chi\chi\to \chi X}}{x^2}\,{\rm d} x\,.
\end{equation}
Using the form of $\langle \sigma v \rangle_{\chi\chi\to \chi X}$ in Eq.~\eqref{eq:semiann_expanded}, the relic abundance in the case of semi-annihilations scales as
\begin{equation}
    \Omega_\chi h^2 \propto  \frac{N_f^2-1}{e_B^2 N_f} \frac{x_f^2}{\sqrt{g_*} }\frac{f_\chi^6}{m_\chi^4}\,{\rm GeV}^{-2}\,.
    \label{eq:OMegah2_app_semi}
\end{equation}
Note that $x_f$ itself depends on the model parameters, as can be seen in Eq.~\eqref{eq:xf_semi-annihilation}. Nevertheless, Eq.\eqref{eq:OMegah2_app_semi} provides a good analytical understanding of how the parameters need to be adjusted to yield the observed relic abundance. In particular, the contours in the $(m_\chi,f_\chi)$ parameter space satisfying $\Omega_\chi h^2=0.12$ follow the scaling relation $f_\chi\propto e_B^{1/3} m_\chi^{2/3}$, as illustrated in Fig.~\ref{fig:semi-ann-full} by the three solid lines corresponding to $e_B=\{0.1,0.1^3,0.1^5\}$.

\subsection{Full Contribution} 
\label{sec:putting_together}

When both semi-annihilation and annihilation processes are included, the freeze-out moment $x_f$ can be found by numerically solving the following equation~\cite{DEramo:2010keq}
\begin{equation}
    x_f=\log\left[0.038c(c+1)\frac{1}{2}\langle\sigma v\rangle_{\chi\chi\rightarrow\chi X}\frac{m_\chi M_{\rm Pl}}{\sqrt{g_{*}x_f}}\right]+\log\left[1+2\frac{(c+2)\langle\sigma v\rangle_{\chi\chi\rightarrow XX}}{(c+1)\langle\sigma v\rangle_{\chi\chi\rightarrow\chi X}}\right]\,.
    \label{eq:xf_full}
\end{equation}
We present an analytic solution for $x_f$ in the non-relativistic limit in App.~\ref{app:Semianalytic}, retaining only the leading-order term in $x^{-1}$ from each cross-section. Moreover, a comparison with numerical results for various benchmarks shows good agreement.

The impact of dark pion annihilations on the freeze-out value $x_f$ appears as a subleading correction in the second term of Eq.~\eqref{eq:xf_full}, on top of the value set by the semi-annihilation process in Eq.~\eqref{eq:xf_semi-annihilation}. This is due to the one-loop suppression of $\langle\sigma v\rangle_{\chi\chi\to XX}$ relative to $\langle\sigma v\rangle_{\chi\chi\to \chi X}$. Therefore, as discussed in \S\ref{sec:annihilation}, annihilations significantly affect the value of $x_f$ only when $m_\chi > 4\pi f_\chi$ and the two thermally-averaged cross-sections become comparable.

The same conclusion pertains for the present-day dark pion energy density. Indeed, Eq.~\eqref{eq:omega_pi_h2} can be generalised to the case where  both processes are kept track of~\cite{DEramo:2010keq}:
\begin{equation}
    \Omega_\chi h^2 = 2\times \frac{1.07\times 10^{9} {\rm GeV^{-1}}}{\sqrt{g_*} M_{\rm Pl}}\left[\int_{x_f}^\infty \frac{\langle\sigma v\rangle_{\chi\chi\to \chi X}+2\langle\sigma v\rangle_{\chi\chi\to XX}}{x^2} \,\text{d}x\right]^{-1}\,,
    \label{eq:omega_pi_h2_full}
\end{equation} 
where again the annihilation contribution becomes relevant only when $\langle\sigma v\rangle_{\chi\chi\to XX}\approx \langle\sigma v\rangle_{\chi\chi\to \chi X}$, \textit{i.e.}, when $m_\chi > 4\pi f_\chi$. 

As it turns out, numerical solutions of the Boltzmann equation, including both semi-annihilation and annihilation contributions, consistently yield values of $m_\chi$ that reproduce the observed relic abundance while satisfying $m_\chi<4\pi f_\chi$. This naturally places us in a regime where semi-annihilations dominate, which explains why the approximation in Eq.~\eqref{eq:OMegah2_app_semi}, derived under the assumption of semi-annihilations only, accurately describes the solid contours in Fig.~\ref{fig:semi-ann-full}. 

\subsection{Parameter Space} 

\begin{figure}[t]
    \centering
    \includegraphics[width=0.95\linewidth]{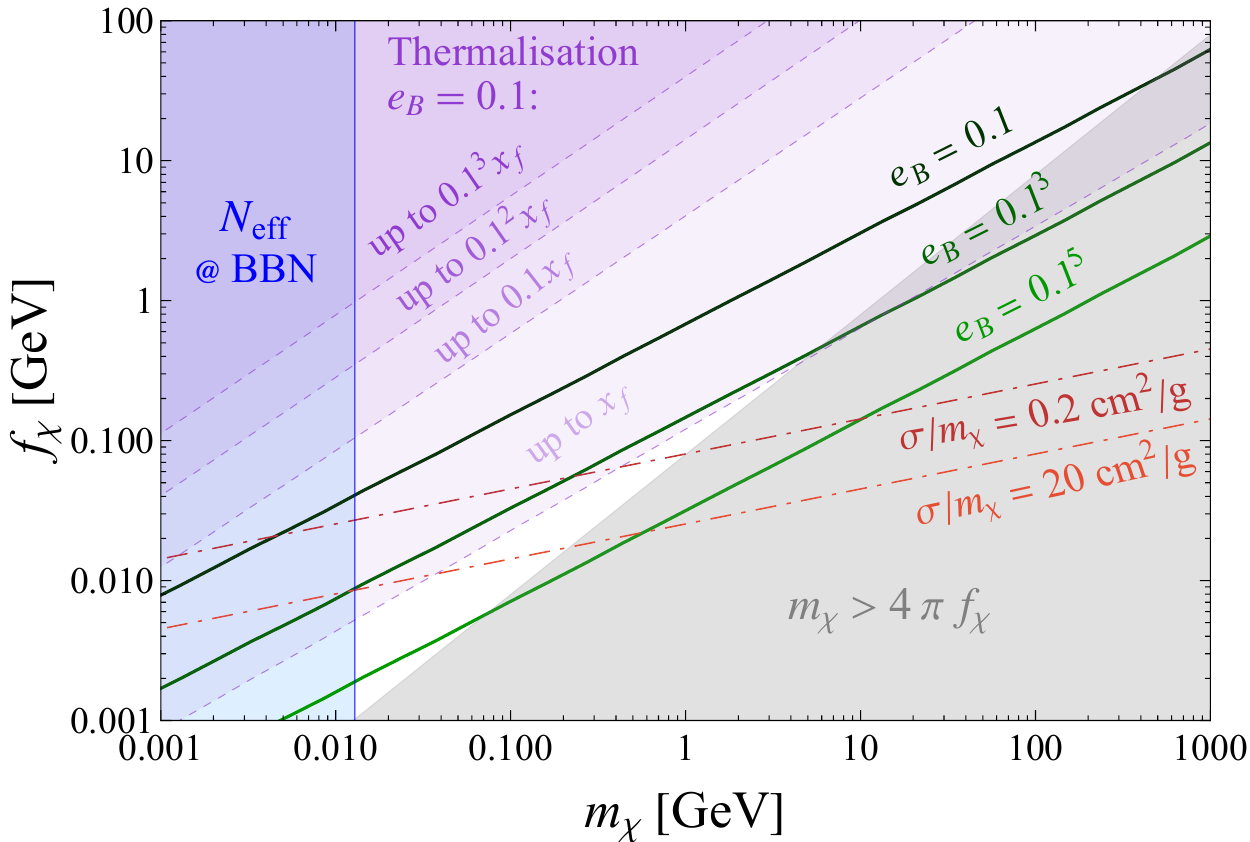}
    \caption{ 
    Solid lines in the $(m_\chi, f_\chi)$ plane indicate the values for which the correct relic abundance of dark pions is achieved: black for $e_B = 0.1$, dark green for $e_B = 0.1^3$, and green for $e_B = 0.1^5$ ($N_f=2$ always). The gray shaded region corresponds to the regime where $m_\chi > 4\pi f_\chi$. The blue region indicates the constraint from $N_{\text{eff}}$ at the time of BBN, excluding values $m_\chi \lesssim 10~\text{MeV}$. Purple regions mark areas of parameter space where thermalisation of dark pions fails to occur by a given point in the cosmic evolution. Different dashed lines represent the boundaries beyond which successful thermalisation is achieved only up to the shown temperatures, expressed in units of $x_f = m_\chi/T_{\rm FO}$, with $T_{\rm FO}$ being the freeze-out temperature. Note, however, that the lack of thermalisation is not a problem because there is no self-heating thanks to the $p$-wave nature of semi-annihilation. Finally, the dash-dotted red lines indicate the constraints from the self-interactions among dark pions.}
    \label{fig:semi-ann-full}
\end{figure}

We are now ready to put things together, summarising all the important constraints in Fig.~\ref{fig:semi-ann-full}.
Since semi-annihilations mainly determine the dark pion relic abundance, the key parameters are $m_\chi$, $f_\chi$, $e_B$, and the dark photon mass $m_X$. 
We find that the interesting dark pion mass range, capable of yielding $\Omega_\chi h^2 = 0.1200 \pm 0.0024$~\cite{Planck:2018vyg},
is 
\begin{equation}
    m_\chi\in[10^{-3},10^{3}] \text{~GeV}, \qquad \text{for~}  e_B=0.1\, ,
\end{equation}
as indicated by the solid dark green contour in Fig.~\ref{fig:semi-ann-full}.
Reducing $e_B$ lowers the required values of $f_\chi$ to achieve the correct relic abundance, which in turn shifts the interesting $m_\chi$ mass range downward -- as shown by the lighter green solid contours. These contours, for $e_B=0.1^3$ and smaller, begin to enter the region where $m_\chi > 4\pi f_\chi$, marked by the gray shaded area in Fig.~\ref{fig:semi-ann-full}, and thus fall outside the EFT range of validity (also the area where annihilations would dominate).

\subsubsection*{Dark Photon Mass Effects}

So far we have restricted to the case in which the dark photon is massless. The effect of introducing a non-zero dark photon mass $m_X$ becomes significant only when it approaches the dark pion mass. Of course, the semi-annihilation channel $\chi \chi\to \chi X$ channel closes in the limit $m_X\to m_\chi$, as can be checked from Eq.~\eqref{eq:sa_mb} in the Appendix, and so we expect that turning on a non-zero $m_X$ for fixed $m_\chi$ requires a larger coupling, {\em i.e.} a reduced value of $f_\chi$, to fit the same relic abundance. Specifically, the required value of $f_\chi$ follows the approximate scaling relation $f_\chi \propto f_{\chi}^{(0)} \Delta^{1/4}$, where $f_{\chi}^{(0)}$ is the value in the massless dark photon limit and $\Delta= (m_\chi-m_X)/m_\chi$ is the mass-splitting. Nevertheless, as long as $\Delta>0.2$, the effect of a nonzero $m_X$ on the final dark pion yield remains below 20\%. Therefore, we adopt the natural assumption of non-degeneracy between the dark pion and photon and set $m_X=0$ throughout.

\subsubsection*{Constraint from BBN}

Another consideration is the limit that $\Delta N_{\it eff}\lesssim 0.4$ at the time of Big-Bang Nucleosynthesis (BBN)~\cite{Nollett:2013pwa}, which leads to a constraint on $m_X$ (and so indirectly on $m_\chi$ too) as follows. If the dark photon is lighter than an MeV and stays in thermal equilibrium at the time of BBN, it would contribute $\Delta N_{\it eff} = \frac{8}{7}$, not allowed by the current limit. We therefore have to assume that the dark photon decays away before the temperature comes down to $T \simeq 2$~MeV, requiring its mass to be above several MeV. Since the dark pion is supposed to be heavier than the dark photon, we conservatively require $m_\chi \gtrsim 10$~MeV, shown as the blue excluded region in Fig.~\ref{fig:semi-ann-full}.

\subsubsection*{Dark Matter Self-Interactions}

As already mentioned at the end of the previous Section, lowering the gauge coupling $e_B$ requires a lower $f_\chi$ for fixed $m_\chi$ in order to keep $\Omega_\chi h^2 \approx 0.12$. However, for $N_f=2$ flavours of dark quark, the chiral Lagrangian predicts self-interactions among the dark pions that become too large for low $f_\chi$ values. To quantify this effect, we report the partial-wave amplitudes $a_{L=0,I=0} = \frac{7m_\chi}{16\pi f_\chi^2}$ and $a_{L=0,I=2} = -\frac{m_\chi}{8\pi f_\chi^2}$ for $\chi \chi \to \chi \chi$ scattering in the isospin channels $I=0,2$. The isospin channel $I=1$ has only a $p$-wave scattering, which can be neglected for our purpose. Among the three dark pions, the flavour-averaged self-interaction cross section is
\begin{equation}
    \sigma^{\rm flavour-avg.}_{\chi \chi\to\chi\chi} = \frac{1}{9} 4\pi a_{L=0,I=2}^2 + \frac{5}{9} 4\pi a_{L=0,I=2}^2 = \frac{23 m_\chi^2}{192\pi f_\chi^4}\ .
    \label{eq:seli_int}
\end{equation} 
Contours of constant $ \sigma^{\rm flavour-avg.}_{\chi \chi\to\chi\chi}$ are shown by dash-dotted red lines in Fig.~\ref{fig:semi-ann-full}. Note, however, that the dark pions may scatter resonantly with a strong velocity dependence~\cite{Chu:2018fzy} through a $\rho$-like~\cite{Tsai:2020vpi} or $\sigma$-like~\cite{Kondo:2022lgg} state, and hence typical upper limits such as $\sigma/m_\chi < 0.2$~cm$^2$/g (see, {\it e.g.}\/, \cite{Ando:2025qtz} for a recent analysis) may not apply.

%%%%%%%%%%%%%%%%%%%%%%%%
\section{Phenomenology}
\label{sec:pheno}
%%%%%%%%%%%%%%%%%%%%%%%%

Having delineated the viable region of parameter space that fits the DM relic abundance, and taking into account other cosmological constraints, we now turn to discuss how this model of DM semi-annihilation via topological freeze-out can be searched for experimentally.

\subsection{Direct and Indirect Detection}

Firstly, and as we have already alluded to, the indirect detection of our dark pion DM via their annihilation $\chi \chi \to XX\to \gamma\gamma$ at the galactic center, halos, or dwarf galaxies, see e.g. \cite{Guo:2023kqt} for bounds on semi-annihilation sceanrios,is suppressed by the $p$-wave nature of the scattering. The constraints are therefore negligible even for the sub-GeV DM mass range. Direct detection is also highly suppressed because the process involves many loops. This is a key feature of our model, and traces back to the topological nature of the $\chi\chi\chi X$ vertex (which lends it a three-derivative suppression).

On the other hand the $X_\mu$ gauge boson, which is kinetically mixed with the photon, can be produced at accelerators and hence allows for interesting phenomenology. We turn our attention to accelerator-based searches below.

\subsection{Dark Photon Bounds}

Recall the bound~\eqref{eq:epsilon} derived above for the dark photon to remain in equilibrium with the SM through the vector portal. Evaluating this at the DM freeze-out temperature gives the following ball-park lower bound on the kinetic mixing parameter $\epsilon$,
\begin{equation} \label{eq:epsilon-2}
    \epsilon \gtrsim 10^{-9} \frac{m_\chi}{\text{GeV}}\ .
\end{equation}
For $m_X$ light enough, we can evade bounds from astrophysics to allow for this kinetic mixing to be large enough: roughly, we need to go down to $m_X \lesssim 10^{-14}$ eV. But, as we have seen, for such light dark photon masses we would contravene the bounds from BBN, and so we focus our interest on the allowed region with heavier dark photon mass, $m_X \gtrsim 1$ GeV. 
Here the experimental bound on the kinetic mixing parameter is very weak compared to~\eqref{eq:epsilon-2}, and comes from various collider measurements -- beam-dump experiments, BaBar, Belle, LEP, and the LHC experiments -- the most relevant of which are discussed below.
We refer the reader to Refs.~\cite{Tsai:2020vpi,Caputo:2021eaa} for compilations of experimental limits.  

\subsection{Collider Phenomenology}

In our model, dark sector states can be produced at collider experiments principally through the kinetic mixing portal -- although there is also the possibility of a Higgs portal that mixes the $U(1)_B$-breaking scalar $\phi$ with the SM Higgs boson {\em viz.} $\mathcal{L} \supset \kappa |H|^2 |\phi|^2$. 

The production of DS particles via the vector portal is suppressed by $\epsilon^2$, while in the case of a Higgs portal the production rates are governed by the Higgs mixing parameter $\kappa$. Here, we mostly focus on the case where the vector portal is the dominant one. Depending on the values of $\epsilon$ and $e_B$ the dark photon can either decay to two dark quarks ($\Gamma_{X\to Q\bar{Q}}\propto e_B^2$) or two leptons ($\Gamma_{X\to l\bar{l}}\propto \epsilon^2$). As we are interested in the case $m_X < m_{\chi}$ and $\chi$ are the lightest particles of the confining sector, two dark quarks are always produced via an off-shell dark photon. Once produced, they will undergo dark sector shower and hadronisation, leading to dark pions in the final state.   

\begin{figure}[t]
\centering
\begin{tikzpicture}
  \begin{feynman}
    \vertex (a);
    \vertex [left=of a] (i) {$e^+$};
    \vertex [below=0.8in of a] (b);
    \vertex [left=of b] (j) {$e^-$};
    \vertex [right=0.8in of a] (c) {$\gamma$};
    \vertex [right=0.6in of b] (d); %b vertex
    \vertex [right=0.4in of d] (x);
    \vertex [above=0.2in of x] (y) {$\bar{Q}_i$};
    \vertex [below=0.2in of x] (z) {$Q_i$};
    \diagram* {
      (i) -- [fermion] (a) -- [fermion] (b) -- [fermion] (j), 
      (a) -- [photon] (c),
      (b) -- [photon,  edge label=$\gamma^* \qquad X$, insertion={[size=4pt,style=thick]0.5}] (d),
      (y) -- [fermion] (d), 
      (d) -- [fermion] (z),
    };
    \vertex [right=1.4in of c] (a1);
    \vertex [left=of a1] (i1) {$e^+$};
    \vertex [below=0.8in of a1] (b1);
    \vertex [left=of b1] (j1) {$e^-$};
    \vertex [right=0.8in of a1] (c1) {$\gamma$};
    \vertex [right=0.6in of b1] (d1); %b vertex
    \vertex [right=0.4in of d1] (x1) {$\chi_b$};
    \vertex [above=0.25in of x1] (y1) {$\chi_a$};
    \vertex [below=0.25in of x1] (z1) {$\chi_c$};
    \vertex [right=1.0in of x1] (w1); % {\footnotesize{[or DS spin-1 $\omega$ resonance]}};
    \diagram* {
      (i1) -- [fermion] (a1) -- [fermion] (b1) -- [fermion] (j1), 
      (a1) -- [photon] (c1),
      (b1) -- [photon,  edge label=$\gamma^* \qquad X$, insertion={[size=4pt,style=thick]0.5}] (d1),
      (d1) -- [scalar] (x1),
      (d1) -- [scalar] (y1),
      (d1) -- [scalar] (z1),
    };
  \end{feynman}
\end{tikzpicture}
\begin{tikzpicture}
  \begin{feynman}
    \vertex (a);
    \vertex [left=of a] (i) {$e^+$};
    \vertex [below=0.8in of a] (b);
    \vertex [left=of b] (j) {$e^-$};
    \vertex [right=0.8in of a] (c) {$\gamma$};
    \vertex [right=0.6in of b] (d); %b vertex
    \vertex [right=0.4in of d] (s);
    \vertex [above=0.1in of s] (t);
    \vertex [below=0.1in of s] (u);
    \vertex [left=0.2in of u] (m0);
    \vertex [left=0.2in of t] (n0);
    \vertex [right=0.2in of u] (m1);
    \vertex [right=0.2in of t] (n1);
    \vertex [right=0.6in of u] (m2);
    \vertex [right=0.6in of t] (n2);
    \vertex [right=1.0in of u] (m3);
    \vertex [right=1.0in of t] (n3);
    \vertex [right=0.8in of u] (x);
    \vertex [right=0.8in of t] (y);
    \vertex [right=1.6in of d] (e);
    \vertex [right=0.4in of e] (x1) {$\chi_b$};
    \vertex [above=0.25in of x1] (y1) {$\chi_a$};
    \vertex [below=0.25in of x1] (z1) {$\chi_c$};
    \diagram* {
      (i) -- [fermion] (a) -- [fermion] (b) -- [fermion] (j), 
      (a) -- [photon] (c),
      (b) -- [photon,  edge label=$\gamma^* \quad\  X^*$, insertion={[size=4pt,style=thick]0.5}] (d),
      (e) -- [fermion, out=135, in=0] (y) -- [fermion,  edge label=$\omega_n$] (t) -- [fermion, out=180, in=45] (d),
      (d) -- [fermion, out=-45, in=180] (u) -- [fermion] (x) -- [fermion, out=0, in=-135] (e),
      (e) -- [scalar] (x1),
      (e) -- [scalar] (y1),
      (e) -- [scalar] (z1),
      (t) -- [gluon] (u),
      (y) -- [gluon] (x),
      (n0) -- [gluon] (m0),
      (n1) -- [gluon] (m1),
      (n2) -- [gluon] (m2),
      (n3) -- [gluon] (m3),
    };
  \end{feynman}
\end{tikzpicture}
\caption{Mono-photon production in $e^+ e^-$ collisions. The cross representations the kinetic mixing vertex $\gamma^\ast \to X$, which carries a suppression by the parameter $\epsilon$. In the régime $s \gg f_\chi$, where $\sqrt{s}$ is the relevant centre-of-mass energy of the collider ($\sim 91$ GeV at a $Z$ factory, or $\sim 11$ GeV at a $B$ factory), the dark photon $b$ decays to a pair of dark quarks. At lower (relative) energy, $s < f_\chi$ or so, the dark photon can decay to a trio of pions via the $\chi \chi \chi X$ semi-annihilation vertex that sets the DM relic abundance in this theory; alternatively, the off-shell dark photon may mix into a spin-1 resonance $\omega_n$ of the dark sector, enabling a `dark spectroscopy' search strategy.\label{fig:mono-photon}
}
\end{figure}
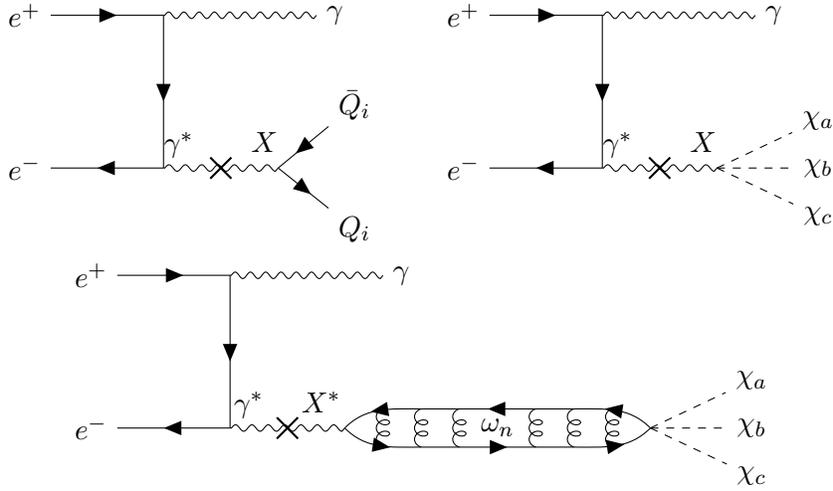

The collider signatures of the model depend on the hierarchy of $e_B$ and $\epsilon$ as well as the lifetime of $X_\mu$. Possible signatures include mono-photon (via the Feynman diagrams shown in Fig.~\ref{fig:mono-photon}), mono-jet, exotic Higgs decays $h\to (\gamma+)\,{\textit{inv.}}$ and four lepton final states, or a dark shower from the decay to dark quarks. 

\subsubsection*{Mono-photon}

For dark photon masses towards the higher end of our preferred mass range, there are important searches with mono-photon final states at the LHC {\it e.g.}\/ in \cite{ATLAS:2020uiq}, where bounds are presented for two different benchmark choices of mediator-lepton, mediator-quark and mediator-dark matter couplings, $\{g_l,\, g_q,\, g_\chi\}$. 
For our model, these bounds can be cast into the $\{m_\chi, m_X\}$
plane, 
excluding the mass range $\{m_\chi, m_X\} \in \{[80,425], [0,840] \}$ GeV for couplings $\{0,\, 0.25,\, 1\}$, and excluding $\{m_\chi, m_X\} \in \{[15,175], [0,350] \}$ GeV for couplings $\{0.01,\, 0.1,\, 1\}$. But note that we cannot directly use these benchmarks because, in our case, both the dark photon coupling to leptons and quarks are proportional to $\epsilon^2$ and cannot be taken independently. Instead, we make use of the limits on the fiducial cross section  $\sigma\times\epsilon_{\rm eff}\times A \lesssim 0.53-2.45$~fb depending on the signal region, with $A$ the acceptance and $\epsilon_{\rm eff}$ the efficiency. Assuming the mono-photon final state occurs from photon pair production, where one of the photons mixes to a dark photon, this sets a limit of $\epsilon\lesssim5\times 10^{-3}-10^{-2}$. The mono-photon final state also arises from s-channel dark photon production via kinetic mixing and an initial-state radiation photon. This process is proportional to $\epsilon^2 \times e_B$ and additionally suppressed due to the dark photon being produced off-shell. A search for a heavy scalar resonance, \textit{e.g.} the $U(1)$ breaking scalar, in the mono-photon final state can be found in \cite{ATLAS:2023eux}. Expected sensitivities for the mono-photon final states at high-luminosity LHC (HL-LHC) are discussed in \cite{ATLAS:2018jsf}, while prospects of this signature at a future muon collider are considered in \cite{Han:2020uak}.

Mono-photon searches have also been performed at B factories, which provide the best constraints for lighter mass dark particles. A search at BaBar excludes $\epsilon \leq 10^{-3}$ for $10^{-3}\text{~GeV~} < m_X < 2 \text{~GeV}$. The bound extends up to $m_X \leq 8$ GeV limiting $\epsilon$ between a few times $10^{-4}$ and $10^{-3}$ \cite{BaBar:2017tiz}. It is expected that Belle II will improve this bound by nearly an order of magnitude to reach $\epsilon \lesssim 3\times 10^{-4}$ \cite{Belle-II:2018jsg}. 

\subsubsection*{Dark Spectroscopy}

In addition, at $e^+ e^-$ colliders such as Belle and BaBar, or indeed at a future high-energy lepton collider such as FCC-ee, kinematic constraints provide more information to pin down invisible decays. Consider $e^+ e^- \rightarrow \gamma X^*$, where the off-shell dark photon $X^*$ mixes into $\omega$-like resonances -- see Fig.~\ref{fig:mono-photon}. Here the energy of the visible photon entirely determines the invariant mass of the invisible system:
\begin{align}
     M_{\rm inv}^2= s \left( 1 - \frac{2E_\gamma}{\sqrt{s}}  \right) .
\end{align}
Therefore, the energy distribution of the mono-photon signal events would exhibit a set of resonant peaks due to $\omega$-resonances in the DS, mirroring states such as $\omega(782)$, $\omega(1420)$, $\omega(1650)$, and $\omega(2220)$ in the case of visible QCD. In addition, the dark sector may well include additional heavier quarks, giving further peaks for resonances analogous to the $J/\psi$ and so on. 
Observing such a spectrum of resonances would, in principle, determine the number of colours, flavours, and quark masses without observing any of the states directly -- clearly a fascinating possibility that was proposed in~\cite{Hochberg:2015vrg,Hochberg:2017khi}.

\subsubsection*{Exotic Higgs Decays}

Higgs decays to visible and dark photons were searched for {\it e.g.}\/ in \cite{CMS:2020krr,ATLAS:2022xlo}. The current limit on this decay is ${\rm{Br}}(h\to\gamma X)\lesssim 0.029$ \cite{ParticleDataGroup:2024cfk}.  Similarly, the bound on invisible Higgs decays is  $h\to \textit{inv.}< 10.7\%$ \cite{ParticleDataGroup:2024cfk}. From $\text{Br}(h\to\gamma X)\sim \text{Br}(h\to\gamma \gamma)\times \epsilon^2$ and $\text{Br}(h\to XX)\sim \text{Br}(h\to\gamma \gamma)\times \epsilon^4$, one can infer bounds on the kinetic mixing parameter of order $\epsilon\lesssim\mathcal{O}(1)$ -- rather weak.

At the HL-LHC, as well as future higgs factories such as FCC-ee and ILC the invisible higgs branching fraction can be probed to $\mathcal{O}(10^{-2}-10^{-3})$ \cite{Peskin:2013xra,Liu:2016zki,Dawson:2022zbb}. With $\text{Br}(h\to XX)\sim \text{Br}(h\to\gamma \gamma)\times \epsilon^4$ this improving the reach on $\epsilon$ in our model by about an order of magnitude. 

\subsubsection*{Mono-jet}

The mono-jet final state originates from $s$-channel dark photon production with an initial-state radiation jet. Both ATLAS and CMS have searched for the mono-jet final state~\cite{ATLAS:2016bek, CMS:2021far}. In \cite{CMS:2021far}, the search was interpreted for a vector mediator with couplings to quarks $g_q (\propto\epsilon) = 0.25$ and couplings to dark matter $g_\chi (=e_B)=1$, excluding $m_\chi\lesssim 115$~GeV for massless dark photons up to $m_\chi\lesssim 300$~GeV for $m_X\sim 575$~GeV. These bounds will loosen for smaller values of $\epsilon$. In addition, for $e_B= 1$ and $m_\chi = m_X/3$ limits on $g_q$ are given ranging from $2\times10^{-2}$ for $m_X =100$~GeV to $0.3$ for $m_X = 2$~TeV. For $m_X < 2m_\chi$ these limits weaken. 

LHC mono-jet limits from LHC have been scaled for HL-LHC projecting $m_\chi\lesssim 200$~GeV exclusion for various dark matter-mediator coupling choices \cite{Boveia:2022jox}.

\subsubsection*{Leptonic Decays}

If $X_\mu$ decays promptly and $\epsilon$ is not too small, we will see leptonic final states. A search for a low-mass resonance decaying to muons sets limits $\epsilon\lesssim2\times 10^{-3}$ for $m_X< 80$GeV and $\epsilon\sim 3\times 10^{-3}-5\times10^{-3}$ for $m_X = 100-200$~GeV~\cite{CMS:2019buh}. Similarly, a search for a low-mass resonance in Higgs decays to four leptons final state sets limits on $\text{Br}(H\to XX)\times \text{Br}(X \to ee\, \text{or}\, \mu\mu)^2\lesssim2\times10^{-6}$ \cite{CMS:2021pcy} leading to $\mathcal{O}(1)$ bounds on $\epsilon$. Assuming a Higgs portal with $\kappa \gg \epsilon$, limits on $\kappa\lesssim 10^{-4}-10^{-3}$ for $m_X \sim 5-60$~GeV were obtained. 

Prospects of searches for leptonically decaying dark photons can be found for HL-LHC and future electron-positron colliders in \cite{Curtin:2014cca,He:2017zzr}. It is expected that these machines can probe $\epsilon\sim\mathcal{O}(10^{-2}-10^{-3})$ via Drell-Yan production of dark photons. If $Br(h\to XX)\sim 0.5\%$ the leptonic decays will be able to probe $\epsilon\sim 10^{-9}-10^{-6}$ at HL-LHC.

\subsubsection*{Semi-visible lepton jets}

Finally, an interesting signature can arise when $X_\mu$ decays promptly to dark quarks, which shower and form dark hadrons, but also emit dark photons as part of the shower process. Thus, part of the shower can become visible when some of the dark photons decay back into SM leptons. The resulting signature consists of a collimated spray of (soft) leptons in association with a large amount of missing energy in the same direction, similar to a semi-visible jet~\cite{Cohen:2015toa,Buschmann:2015awa}.

\section{Conclusion and Outlook}
\label{sec:conclusion}

In this paper we have shown that a simple QCD-like model of the dark sector allows for dark pion dark matter that freezes out via semi-annihilation into a dark photon. This is possible due to the existence of a topological interaction between the dark pions $\chi$ and the $U(1)_B$ gauge boson $X_\mu$, via the Skyrme current, that is produced by gauging baryon number in the ultraviolet. The model is technically natural, consistent with a range of dark matter mass $10~{\rm MeV} \lesssim m_\chi \lesssim 1~{\rm TeV}$, and evades the limits from indirect detection due to the $p$-wave nature of the semi-annihilation. 

The model has rich phenomenology in cosmology and laboratory experiments. Towards the lighter limit of the allowed mass range, the semi-annihilation process may lead to 
significant self-interactions that can modify the galactic dynamics at small scales. 
We discussed potential signatures at colliders, such as monojets at the LHC, invisible Higgs decays at the LHC and at a future Higgs factory, dark spectroscopy at SuperKEKB, and more. We find it intriguing that such a simple model gives rise to a wide ray of rich phenomenology in a way fully consistent with current observational and experimental constraints.

In addition to pursuing more comprehensive phenomenological studies, there are several other future directions we plan to explore. 
Firstly, it is known that the $2\to 1$ process $\chi \chi \to \chi X$ at the heart of this work can give rise to an alternative DM thermal history besides the freeze-out option explored here, in which the DM abundance starts off very small and is rapidly generated through $X\chi \to \chi\chi$. This scenario is known as {\em explosive freeze-in} \cite{Bringmann:2021tjr}, and it would be interesting to characterise the viable parameter space for this option. 
Secondly, our restriction to $N_f=2$ flavours of dark quark can be generalised to $N_f >2$; in this case, the model also features a WZW term which induces $3\to 2$ number changing processes, as has been exploited in theories of Strongly Interacting Massive Particle (SIMP) DM~\cite{Hochberg:2014dra,Hochberg:2014kqa}.
Including both this WZW term and the gauged baryon number interaction we introduced in this paper, it would be intriguing to study the interplay between the $3\chi \to 2\chi$ WZW-induced process, which depends on the decay constant $f_\chi$, and the $2\chi \to \chi X$ semi-annihilation channel, which depends not only on $f_\chi$ but also on the gauge coupling $e_B$.

\section*{Acknowledgements}
We thank Eric Kuflik, Jesse Thaler, Ayuki Kamada, and Mathias Becker for useful discussions.
The work of H.\,M. is supported by the Director, Office of Science, Office of High Energy Physics of the U.S. Department of Energy under the Contract No. DE-AC02-05CH11231, by the NSF grant PHY-2210390, by the JSPS Grant-in-Aid for Scientific Research JP23K03382, MEXT Grant-in-Aid for Transformative Research Areas (A) JP20H05850, JP20A203, Hamamatsu Photonics, K.K, and Tokyo Dome Corportation. In addition, HM is supported by the World Premier International Research Center Initiative (WPI) MEXT, Japan. Likewise, H.\,M. thanks the CERN theory group for their warm hospitality during the \emph{Crossroads between Theory and Phenomenology} Program, where this work was initiated. The work of N.\,S. has received funding from the INFN Iniziative Specifica APINE. C.\,S. is supported by the Office
of High Energy Physics of the U.S. Department of Energy under contract DE-AC02-05CH11231.

\appendix

%%%%%%%%%%%%%%%%%%%%%%%%%%%%%%
%%%%%%%%%%%%%%%%%%%%%%%%%%%%%%
\section{Semi-Annihilation Computations}
\label{app:semi_annihilation}
%%%%%%%%%%%%%%%%%%%%%%%%%%%%%%
%%%%%%%%%%%%%%%%%%%%%%%%%%%%%%

In this appendix, we provide details regarding the computation of the thermally-averaged cross-section for the semi-annihilation process. 
The corresponding matrix-squared element for the process $\chi^a(p_1)\chi^b(p_2) \to \chi^c(p_3)X^\mu(p_4)$, summed over the final state polarisations and flavours, and averaged over the initial state flavours, is computed using the Feynman rule in Eq.~\eqref{eq:semiannihilation} and reads
\begin{equation}
    \Sigma|\overline{\mathcal{M}}|^2 = \frac{N_f}{N_f^2-1}\frac{e_B^2}{64\pi^4 f_\chi^6}(s-4m_\chi^2)\left(s-(m_\chi+m_X)^2\right)\left(s-(m_\chi-m_X)^2\right)\sin^2\theta\,,
    \label{eq:semi_M_sq_avg}
\end{equation}
with the Mandelstam variable $s$ defined as $s = (p_1+p_2)^2$. In the center-of-mass frame, we have the following momenta configuration when the incoming pions collide along the $z$-direction and the outgoing photon makes an angle $\theta$ with the $z$-axis in the $y-z$ plane
\begin{align}
    p_1 &= \left(\frac{\sqrt{s}}{2},0,0,\frac{\sqrt{s-4m_\chi^2}}{2}\right)\,,\\
    p_2 &= \left(\frac{\sqrt{s}}{2},0,0,-\frac{\sqrt{s-4m_\chi^2}}{2}\right)\,,\\
    p_3 &= \left(\frac{s+m_\chi^2-m_X^2}{2\sqrt{s}},0,|\vec{p}_f|\sin\theta,|\vec{p}_f|\cos\theta\right)\,,\\
    p_4 &= \left(\frac{s-m_\chi^2+m_X^2}{2\sqrt{s}},0,-|\vec{p}_f|\sin\theta,-|\vec{p}_f|\cos\theta\right)\,,
\end{align}
where 
\begin{equation}
    |\vec{p}_f|=\sqrt{\frac{m_X^4+(s-m_\chi^2)^2-2m_X^2(s+m_\chi^2)}{4s}}\,.
\end{equation}
The differential cross-section reads
\begin{equation}
    \frac{d\sigma_{\chi\chi\to\chi X}}{d\Omega} = \frac{1}{64\pi^2 s} \frac{|\vec{p}_f|}{|\vec{p}_i|} \Sigma|\overline{\mathcal{M}}|^2\,,
    \label{eq:semi_xsec_diff}
\end{equation}
with $|\vec{p}_i| =\sqrt{s-4m_\chi^2}/2$. Now, it is possible to use Eqs.~\eqref{eq:semi_M_sq_avg}--\eqref{eq:semi_xsec_diff} and perform the $\theta$-integration with $d\Omega=2\pi\sin\theta d\theta$ to arrive to
\begin{equation}
    \sigma_{\chi\chi\to\chi X} = \frac{N_f}{N_f^2-1}\frac{e_B^2}{1536\pi^5} \frac{(m_X^4+(s-m_\chi^2)^2-2m_X^2(s+m_\chi^2))^{3/2}}{f_\chi^6 s^{3/2}}\,,
    \label{eq:sa_mb}
\end{equation}
which, in the limit $m_X\to 0$
\begin{equation}
    \sigma_{\chi\chi\to\chi X} = \frac{N_f}{N_f^2-1}\frac{e_B^2}{1536\pi^5} \frac{s^2}{f_\chi^6} \left(1-\frac{m_\chi^2}{s}\right)^3 \left(1-\frac{4m_\chi^2}{s}\right)^{1/2}\,,
\end{equation}
reproduces Eq.~\eqref{eq:semi_xsec} from the main text.
Following Refs.~\cite{Gondolo:1990dk,Edsjo:1997bg}, we compute the thermally averaged cross section times the M{\o}ller velocity, $v=\sqrt{|\mathbf{v}_1-\mathbf{v}_2|^2 - |\mathbf{v}_1\times\mathbf{v}_2|^2}$, as the following integral
\begin{equation}
		\langle \sigma v
		\rangle_{\chi\chi\to\chi X} = \frac{\int_{4m_\chi^2}^{\infty}\sigma_{\chi\chi\to\chi X}\sqrt{s}(s-4m_\chi^2)\,{\rm{K}}_1(\sqrt{s}/T)\,{\rm{d}}s}{8m_\chi^4 T\,{\rm K}_2^2(m_\chi/T)}\,,
		\label{eq:thavgxsec}
\end{equation}
with ${\rm K}_i$ being the modified Bessel functions of the $i$-th order. Solving the integral, we find
\begin{align}
    \langle \sigma v \rangle_{\chi\chi\to\chi X} &= \frac{N_f}{N_f^2-1} \frac{e_B^2}{4096 \pi^{9/2}}\frac{m_\chi^4}{f_\chi^6} \frac{x}{{\rm K}_2^2(x)} \left[64\,{\rm G}^{3,0}_{1,3}\left(\genfrac{}{}{0pt}{}{-2}{-\frac{9}{2}\,-\frac{1}{2}\,\frac{1}{2}}|\,x^2\right)-48\,{\rm G}^{3,0}_{1,3}\left(\genfrac{}{}{0pt}{}{-1}{-\frac{7}{2}\,-\frac{1}{2}\,\frac{1}{2}}|\,x^2\right)\right.\nonumber\\
    &\left.+12\,{\rm G}^{3,0}_{1,3}\left(\genfrac{}{}{0pt}{}{0}{-\frac{5}{2}\,-\frac{1}{2}\,\frac{1}{2}}|\,x^2\right)-{\rm G}^{3,0}_{1,3}\left(\genfrac{}{}{0pt}{}{1}{-\frac{3}{2}\,-\frac{1}{2}\,\frac{1}{2}}|\,x^2\right)\right]\,,
    \label{eq:semi_thermal_avg_xsec}
\end{align}
where $x=m_\chi/T$, and ${\rm G}(x^2)$ is the Meijer G-function~\cite{ADAMCHIK1995283}.\\

\noindent The expression for $\langle \sigma v
		\rangle_{\chi\chi\to\chi X}$ can be expanded in the inverse powers of $x=m_\chi/T$. Up to $\mathcal{O}(x^{-3})$, the corresponding expression reads
\begin{equation}
    \langle \sigma v
		\rangle_{\chi\chi\to\chi X} = \frac{N_f}{N_f^2-1} \frac{e_B^2}{384 \pi^{5}}\frac{m_\chi^4}{f_\chi^6} \left(\frac{81}{16} \,x^{-1} + \frac{891}{32}\,x^{-2} -\frac{66825}{512}\,x^{-3}+\dots\right)\,.
        \label{eq:semi_ann_expanded}
\end{equation}
Furthermore, it correctly reproduces the result in the non-relativistic limit, $x\gg 1$. Indeed, starting from Eq.~\eqref{eq:semi_xsec}, the non-relativistic limit of the thermally averaged cross-section is
\begin{align}
    \langle \sigma v
		\rangle_{\chi\chi\to\chi X}^{\rm nr}&= \biggl\langle\sigma_{\chi\chi\to\chi X} \cdot 2 \sqrt{1-4m_\chi^2/s} \biggr\rangle\,,
        \label{eq:semi_nonrel}
\end{align}
with $s=4m_\chi^2 + m_\chi^2 v^2$. Therefore, expanding the right-hand side of Eq.~\eqref{eq:semi_nonrel} in $v^2$ and using 
\begin{equation}
    \langle v^n\rangle = \frac{2^{(n-1)/2}\Gamma(\frac{n+3}{2})}{\Gamma(\frac{3}{2})}\, x^{-n/2}\,,
\end{equation}
we find
\begin{align}
    \langle \sigma v
		\rangle_{\chi\chi\to\chi X}^{\rm nr}&= \frac{N_f}{N_f^2-1} \frac{e_B^2}{384 \pi^{5}}\frac{m_\chi^4}{f_\chi^6} \left(\frac{81}{16} \,x^{-1} + \frac{405}{128}\,x^{-2} + \frac{945}{2048}\,x^{-3}+\dots\right)\,,
        \label{eq:sigmav_nr}
\end{align}
with the leading term matching the one in Eq.~\eqref{eq:semi_ann_expanded}.

%%%%%%%%%%%%%%%%%%%%%%%%%%%%%%
%%%%%%%%%%%%%%%%%%%%%%%%%%%%%%
\section{Annihilation Computations}
\label{app:annihilation}
%%%%%%%%%%%%%%%%%%%%%%%%%%%%%%
%%%%%%%%%%%%%%%%%%%%%%%%%%%%%%

The operators in the chiral Lagrangian can mediate the $\chi^a(p_1)\chi^b(p_2) \to X^\mu(p_3) X^\nu(p_4)$ process. There are two independent dimension-six operators, as given also in the main text, 
\begin{align}
    \mathcal{L}_{\chi{\rm PT}} \supset \left(\frac{e_B}{16\pi^2 f_\chi}\right)^2 \left(\lambda_1 \left(\partial_\alpha U^\dagger\right) \left(\partial^\alpha U\right) X_{\mu\nu}\,X^{\mu\nu}+\lambda_2 \left(\partial_\alpha U^\dagger\right) \left(\partial^\nu U\right) X_{\mu\nu}\,X^{\mu\alpha}\right)\,,
\end{align}
where $\lambda_1$ and $\lambda_2$ are $\mathcal{O}(1)$ Wilson coefficients (in general complex), and the overall size is set by the dimensionful factor in front. The corresponding Feynman rules read
\begin{align}
    \mathcal{O}_1\,\,&:\,\, -8i\lambda_1 \,p_1.p_2\left(p_4^\mu p_3^\nu - p_3.p_4\, g^{\mu\nu}\right)\delta^{ab}\,,\label{eq:O1}\\
    \mathcal{O}_2\,\,&:\,\,2i\lambda_2 \left(p_1\cdot p_4 \,p_2^\mu p_3^\nu + p_1\cdot p_3 \, p_4^\mu p_2^\nu -p_3\cdot p_4\left(p_1^\mu p_2^\nu + p_2^\mu p_1^\nu \right)\right.\nonumber\\
    & \left.\quad+ p_2\cdot p_4\left(p_1^\mu p_3^\nu - p_1\cdot p_3\, g^{\mu\nu}\right)  + p_2\cdot p_3\left(p_4^\mu p_1^\nu - p_1\cdot p_4 \,g^{\mu\nu}\right)\right)\delta^{ab}\,,\label{eq:O2}
\end{align}
multiplied by the overall size of the UV contribution, $(e_B/16\pi^2 f_\chi)^2$.

We can use this to compute the thermally averaged cross section as outlined in Sec.~\ref{sec:semi_annihilation}. The result for the averaged matrix element squared reads
\begin{align}
    \sum |\overline{\mathcal{M}}|^2& = \frac{1}{N_f^2-1}\left(\frac{e_B}{16\pi^2 f_\chi^2}\right)^4 \bigg[4s^2(s-2m_\chi^2)^2\left({\rm Re}(\lambda_1\,\lambda_2^*) + 2 |\lambda_1|^2\right)\\
    &+ \left(2m_\chi^4 s^2 + \frac{1}{2}s^2 (s-2m_\chi^2)^2-4m_\chi^2 s (t-m_\chi^2)(u-m_\chi^2)+2(t-m_\chi^2)^2(u-m_\chi^2)^2\right)|\lambda_2|^2\bigg]\,.\nonumber
\end{align}
The corresponding cross-section is
\begin{align}
    \sigma_{\chi\chi\to XX} &= \frac{1}{N_f^2-1}\left(\frac{e_B}{16\pi^2 f_\chi^2}\right)^4\frac{1}{8\pi}\, s\sqrt{\frac{s}{s-4m_\chi^2}}\\ 
    &\times\left[(s-2m_\chi^2)^2\left({\rm Re}(\lambda_1\,\lambda_2^*)+ 2 |\lambda_1|^2\right)
    +\frac{1}{120} (92 m_\chi^4 - 76 m_\chi^2 s + 17 s^2) |\lambda_2|^2\right]\,,\nonumber
\end{align}
and the thermally averaged cross-section 
\begin{align}
    \langle \sigma v
		\rangle_{\chi\chi\to XX} &= \frac{1}{N_f^2-1}\left(\frac{e_B}{8\pi^2 f_\chi^2}\right)^4\frac{16 m_\chi^6}{\sqrt{\pi}} \frac{x}{{\rm K}_2^2(x)}\nonumber\\
    &\left[{\rm G}^{3,0}_{1,3}\left(\genfrac{}{}{0pt}{}{-4}{-\frac{11}{2}\,-\frac{1}{2}\,\frac{1}{2}}|\,x^2\right)\left({\rm Re}(\lambda_1\,\lambda_2^*) + 2|\lambda_1|^2+\frac{17}{2}|\lambda_2|^2\right)\right.\nonumber\\
    &-{\rm G}^{3,0}_{1,3}\left(\genfrac{}{}{0pt}{}{-3}{-\frac{9}{2}\,-\frac{1}{2}\,\frac{1}{2}}|\,x^2\right)\left({\rm Re}(\lambda_1\,\lambda_2^*) + 2|\lambda_1|^2+\frac{19}{2}|\lambda_2|^2\right)\nonumber\\
    &\left.+\frac{1}{4}{\rm G}^{3,0}_{1,3}\left(\genfrac{}{}{0pt}{}{-2}{-\frac{7}{2}\,-\frac{1}{2}\,\frac{1}{2}}|\,x^2\right)\left({\rm Re}(\lambda_1\,\lambda_2^*) + 2|\lambda_1|^2+\frac{23}{2}|\lambda_2|^2\right)\right]\,.
    \label{eq:ann_thermal_avg_xsec}
\end{align}
    Expanding $\langle \sigma v\rangle_{\chi\chi\to XX}$ in the limit of large $x$, we obtain
  \begin{align}
        \langle \sigma v
		\rangle_{\chi\chi\to XX} &\approx \frac{1}{N_f^2-1}\left(\frac{e_B}{16\pi^2 f_\chi^2}\right)^4\frac{m_\chi^6}{2\pi} |4\lambda_1+\lambda_2|^2 \left(1+\frac{15}{8}x^{-1}+\mathcal{O}(x^{-2})\right)\,.
        \label{eq:sigmav_ann_nr}
    \end{align}

%%%%%%%%%%%%%%%%%%%%%%%%%%%%%%
%%%%%%%%%%%%%%%%%%%%%%%%%%%%%%
\section{Semi-Analytical {\em vs.} Numerical Results}
\label{app:Semianalytic}
%%%%%%%%%%%%%%%%%%%%%%%%%%%%%%
%%%%%%%%%%%%%%%%%%%%%%%%%%%%%%
In this appendix, we compare the numerical results for $x_f$ (the freeze-out moment) and $Y_\chi(x_{\infty})$ (the asymptotic limit of the dark pion yield today, $x\to x_\infty$) with the semi-analytic solutions that we present below.

We first solve for $x_f$ in Eq~\eqref{eq:xf_full} in the non-relativistic limit, keeping only the leading order in $x^{-1}$ from each cross-section. For later convenience, we write the cross-sections as $\langle\sigma v\rangle_{\chi\chi\rightarrow XX}\approx\sigma _a x^{-n_a}$ and $\langle\sigma v\rangle_{\chi\chi\rightarrow \chi X}\approx\sigma _s x^{-n_s}$, where $n_i$ denotes the leading power in the $x^{-1}$ expansion ($n_a=0$ for $s$-wave annihilations, and $n_s=1$ for $p$-wave semi-annihilations). Therefore, we have
\begin{align}
    x_f&\simeq\log\left[c(c+1)0.0382\frac{1}{2}\frac{M_{\rm Pl}m_\chi\sigma_s}{\sqrt{g_*}}\right]-\left[n_s+\frac{1}{2}\right]\log[x_f]+\log\left[1+2\frac{\sigma_a}{\sigma_s}\frac{c+2}{c+1}x_f\right]\,.
\end{align}
For annihilation processes, the best fit for $c$ is to choose $c(c+2)=n_a+1$ ~\cite{Kolb_Turner_2018}. Since semi-annilation is the dominant process here, we choose $c(c+1)=n_s+1$. For our $p$-wave process $c=1$.
\begin{align}
    x_f&\simeq\log\left[c(c+1)0.0382\frac{1}{2}\frac{M_{\rm Pl}m_\chi\sigma_s}{\sqrt{g_*}}\right]-\left[c(c+1)-\frac{1}{2}\right]\log[x_f]+\log\left[1+2\frac{\sigma_a}{\sigma_s}\frac{c+2}{c+1}x_f\right]\,.
\end{align}
Moreover, we can use that $\frac{\sigma_a}{\sigma_s}\sim \frac{2}{3(16\pi^2)^2}\frac{m_\chi^2 e_B^2}{f_\chi^2 N_f}$, with $x_f$ of order 10, and for the choices of $m_\chi$ and $f_\chi$ considered, the last $\log$ in the expression above is subleading. In this case, we find
 \begin{equation}  
\begin{aligned}
     x_f&\simeq  \ln\left[0.038 c(c+1)\frac{1}{2g_*^{1/2}}M_{\rm Pl}m_\chi\sigma_s\right]-\left(c(c+1)-\frac{1}{2}\right)\ln\left\{ \ln\left[0.038 c(c+1)\frac{1}{2g_*^{1/2}}M_{\rm Pl}m_\chi\sigma_s\right]\right\}\\
    &+\ln\left\{1+2\frac{\sigma_a}{\sigma_s }\frac{(c+2)}{(c+1)}\ln\left(0.038\frac{1}{2g_*^{1/2}}M_{\rm Pl}m_\chi\sigma_s\right)\right\}\,,
\end{aligned}
\end{equation}
and
\begin{align}
    Y_{\infty}&\simeq\frac{7.8\, g_*^{-1/2}}{M_{\rm Pl}m_\chi (2\sigma_a x_f^{-1}+\frac{1}{2}\sigma_sx_f^{-2})}\,.
\end{align}
The results and their comparison to the fully numerical solutions for a couple of benchmark values of the parameters are shown in the tables below. \\

$\mathbf{N_f=2,e_B=1,\lambda_1=1,\lambda_2=1,c=1}:$
\begin{center}
\begin{tabular}{ |c|c|c|c|c|c|c| } 
 \hline
 $f_\chi$ [GeV] & $m_\chi$ [GeV] & $g_*$& $x_f$& $Y_{\infty}$ & Y(50) (numeric)& Y(75) (numeric)\\ 
 \hline
 5 & 2.8 &247/4 & 19.55 & $9.55\cdot 10^{-11}$ & $1.14\cdot 10^{-10} $ & $1.04\cdot 10^{-10}$\\ 
 50 & 94 &303/4 & 23.01 & $ 2.77\cdot 10^{-12}$ & $3.55\cdot 10^{-12}$ & $3.09\cdot 10^{-12}$ \\ 
 500&3175 &345/4 &26.66 & $6.91  \cdot 10^{-14}$ & $1.01\cdot 10^{-13}$ &$8.18\cdot 10^{-14}$\\
 \hline
\end{tabular}
\end{center}

$\mathbf{N_f=2,e_B=0.01,\lambda_1=1,\lambda_2=1,c=1:}$
\begin{center}
\begin{tabular}{ |c|c|c|c|c|c|c| } 
 \hline
 $f_\chi$ [GeV] & $m_\chi$ [GeV] & $g_*$& $x_f$& $Y_{\infty}$ & Y(50) (numeric)& Y(75) (numeric)\\ 
 \hline
 5 & 2.8 &247/4 & 11.06 & $3.06\cdot 10^{-7}$ & $3.33\cdot 10^{-7} $ & $3.23\cdot 10^{-7}$\\ 
 50 & 94 &303/4 & 14.38 & $1.09\cdot 10^{-8}$ & $1.22\cdot 10^{-8}$ & $1.16\cdot 10^{-8}$ \\ 
 500&3175 &345/4 &17.83 & $3.59  \cdot 10^{-10}$ & $4.18\cdot 10^{-10}$ &$3.86\cdot 10^{-10}$\\
 \hline
\end{tabular}
\end{center}
\bibliographystyle{utcaps_mod}
\bibliography{Refs}

\end{document}